\documentclass[pra,twocolumn,10pt,aps,superscriptaddress,longbibliography]{revtex4-1}  
\setcitestyle{square}
\usepackage{graphicx}  
\usepackage{float}
\usepackage{dcolumn}   
\usepackage{bm}        
\usepackage{amssymb}   
\usepackage{amsmath}
\usepackage{isomath}
\usepackage[usenames, dvipsnames]{color}
\usepackage{soul}
\usepackage{physics}
\usepackage[normalem]{ulem}
\usepackage{xpatch}
\usepackage{IEEEtrantools}
\usepackage{lipsum}

\usepackage{xcolor}
\usepackage{hyperref}
\hypersetup{
     colorlinks   = true,
     linkcolor    = red,
     citecolor    = cyan,
     urlcolor     = magenta
     }
\usepackage{tikz}

\begin{document}

\widetext
\title{Stability of topological edge states under strong nonlinear effects}

\author{Rajesh Chaunsali}
\email{rajeshcuw@gmail.com}
\affiliation{LAUM, CNRS, Le Mans Universit\'{e}, Avenue Olivier Messiaen, 72085 Le Mans, France}

\author{Haitao Xu}
\affiliation{Center for Mathematical Science, Huazhong University of Science and Technology, Wuhan, Hubei 430074, PR China}

\author{Jinkyu Yang}
\affiliation{Aeronautics and Astronautics, University of Washington, Seattle, WA 98195-2400, USA}

\author{Panayotis G. Kevrekidis}
\email{kevrekid@umass.edu}
\affiliation{Department of Mathematics and Statistics, University of Massachusetts, Amherst, MA 01003-4515, USA}
\affiliation{Mathematical Institute, University of Oxford, OX26GG, UK}

\author{Georgios Theocharis}
\email{georgiostheocharis@gmail.com}
\affiliation{LAUM, CNRS, Le Mans Universit\'{e}, Avenue Olivier Messiaen, 72085 Le Mans, France}

\date{\today} 
             
\begin{abstract}
We examine the role of strong nonlinearity on the topologically-robust edge state in a one-dimensional system. 
We consider a chain inspired from the Su-Schrieffer–Heeger model, but with a finite-frequency edge state and the dynamics governed by second-order differential equations. 
We introduce a \textit{cubic} onsite-nonlinearity and study this nonlinear effect on the edge state's frequency and linear stability. 
Nonlinear continuation reveals that the edge state loses its typical shape enforced by the chiral symmetry and becomes generally unstable due to various types of instabilities that we analyze using a combination of spectral stability and Krein signature analysis. 
This results in an initially-excited nonlinear-edge state shedding its energy into the bulk over a long time. However, the stability trends differ both qualitatively and quantitatively when softening and stiffening types of nonlinearity are considered.  
In the latter, we find a frequency regime where nonlinear edge states can be linearly stable. This enables high-amplitude edge states to remain spatially localized  without shedding their energy, a feature that we have confirmed via long-time dynamical simulations.
Finally, we examine the robustness of frequency and stability of nonlinear edge states against disorder, and find that those are more robust under a chiral disorder compared to a non-chiral disorder. 
Moreover,  the frequency-regime where high-amplitude edge states were found to be linearly stable remains intact in the presence of small amount of disorder of both types. 


\end{abstract}

\maketitle

\section{Introduction}
Band topology has emerged as a mathematical tool in understanding fundamental properties of electronic materials \citep{Hasan2010}. 
It has also led to exciting developments in the bosonic systems, such as cold atom lattices \cite{Cooper2019}, photonics \cite{Ozawa2019} and phononics \cite{Susstrunk2016, Ma2019}.
The main idea of this notion of topology is to characterize the dispersion properties of an infinite (bulk) material and predict how the boundaries of its finite counterpart behave \cite{Bernevig2013}.
This, so-called ``bulk-boundary correspondence'', has turned out to be a direct route to design systems with interesting energy localization properties on their corners, edges, and surfaces \cite{Kane2005, Wan2011, Benalcazar2017}. 
The topological nature of the bulk also imparts certain robustness to the boundary properties, and therefore, those become insensitive to imperfections, i.e., ``topologically robust". 

Though this framework is powerful in predicting and designing exotic systems in various spatial and synthetic dimensions, it is commonly linked to linear dynamics. Therefore, one of the emerging questions in the field of topological materials is: How does nonlinearity affect the characteristics of a topological system? This includes not only studying the effect of nonlinearity on the topologically-robust properties, but also exploring ways to predict purely nonlinear states. 
Recent studies on the interplay between topology and nonlinearity have sparked a tremendous interest along these lines \cite{Smirnova2020}. 
For example, nonlinearity has been used as the tuning knob to modulate the frequency and generate the harmonics of edge states~\cite{Dobrykh2018, Pal2018, Vila2019, Kruk2019, Wang2019, Darabi2019, Zhou2020}. It has also been used to make topologically-robust solitons propagating on edges~\citep{Ablowitz2014, Leykam2016, Kartashov2016, Snee2019, Tao2020,Mukherjee2020b}. 
Furthermore, studies have shown that insights from topological band theory can help us interpret nonlinear solutions, such as gap solitons~\citep{Lumer2013, Solnyshkov2017, Smirnova2019, Marzuola2019, Mukherjee2020}, nonlinear Dirac cones \cite{Bomantara2017}, ``self-induced'' boundary states~\citep{Leykam2016, Hadad2016, Hadad2018a, Savelev2018, Chaunsali2019, Zangeneh2019} and domain walls~\citep{Chen2014, Hadad2017, Poddubny2018}. 
In a driven-damped system, the chaotic dynamics have been shown to exhibit topological features \cite{Engelhardt2017}.
Recently, stability of topological states in periodically driven systems such as nonlinear quantum walks is also discussed \cite{Gerasimenko2016, Bisianov2019, Mochizuki2020}.

While most of the previous works have explored weakly nonlinear regimes where nonlinearity is considered as a perturbation to the topological states, the works in strongly nonlinear regime have been relatively scarce and restricted to only specific setups. 
In particular, the study of stability of topological states in these regimes is a quite subtle issue that requires a systematic approach to investigate the role of several possible types of nonlinearity across different platforms \citep{Lumer2016, Shi2017}. In addition, previous works on the stability of topological states typically involve dynamics that is governed by  first-order ordinary differential equations (ODEs), while the dynamics of second-order ODEs, for example, in phononics \cite{Pal2018} and electrical circuits \cite{Palmero2020} are less explored.

Here we explore the linear stability of topological edge states in the strongly nonlinear regime of a mechanical lattice. We are particularly interested in mechanical systems since they can host a variety of nonlinear functional forms \cite{Theocharis2013, Fraternali2015, Yasuda2019, Deng2020}, and thus, provide a versatile platform to study rich physics resulting from the interplay between topology and nonlinearity.  Moreover, it is possible to use shape optimization algorithms to identify structures that provide specified nonlinear mechanical responses~\cite{Jutte2008, Wang2014, Clausen2015} that could be required to enhance the desired characteristics of the topological states. From a more practical point of view, such mechanical systems promise  efficient solutions for applications such as vibration isolation, sensing, noise mitigation, and energy harvesting \cite{Hussein2014, Cummer2016, Bertoldi2017}.

In this study, we take inspiration from the Su-Schrieffer–Heeger (SSH) model~\citep{Su1979} -- one of the most foundational models for band topology -- and construct a spring-mass lattice with an onsite nonlinearity. Importantly, this supports a band gap, along with an edge state, centered at a \textit{finite} frequency in the linear limit~\citep{Prodan2009, Susstrunk2016, Chaunsali2017}. We choose the cubic form of nonlinearity for simplicity. Also, with the goal of demonstrating how the type of nonlinearity can drastically change the dynamics in a nonlinear regime, we consider both \textit{softening} and \textit{stiffening} types of nonlinearity and compare their effects on the topological edge state. 
We closely examine the amplitude and phase of its Floquet multipliers (FMs), which characterize the linear stability of the corresponding periodic orbit. We employ a Krein signature analysis \cite{Marin1998, Aubry2006, Flach2008} to determine the regions and nature of instabilities.
A key finding is that stiffening nonlinearities may allow for regimes of linear stability of the nonlinear topological states identified herein.
Finally, we introduce chiral and non-chiral types of disorder in the chain and study their effect on the nonlinear continuation and stability for both types of nonlinearity. We observe that chiral disorder leads to lesser variations in nonlinear continuation and stability in general. Moreover, in the presence of weak disorder of both types in the stiffening system, it is possible to retain the frequency regime where nonlinear edge states were found to be linearly stable.

\section{System and the topological edge state} 

 \begin{figure}[!h]
\centering
\includegraphics[width=\columnwidth]{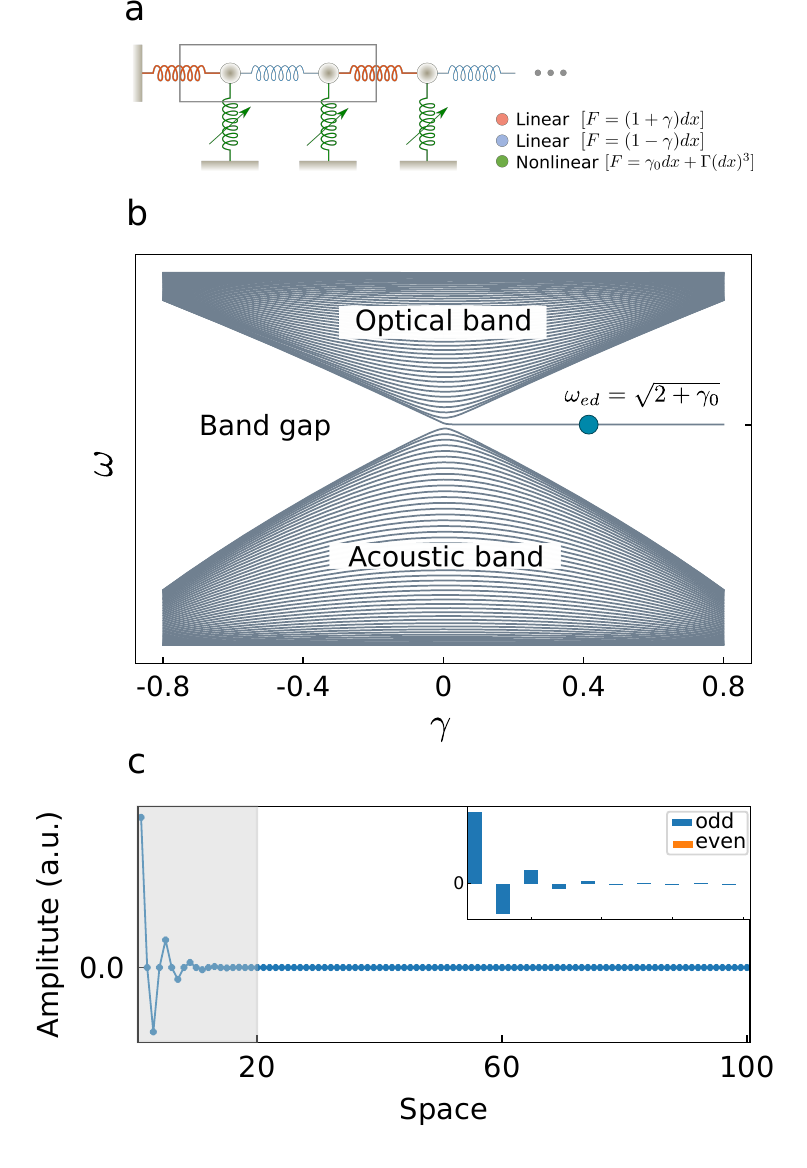}
\caption{(a) A chain of masses that are interconnected with linear springs and grounded with nonlinear springs. (b) Spectrum of the linearized chain ($\Gamma=0$) as a function of $\gamma$. For $\gamma>0$ we see a topological edge state emerging at $\omega_{ed} = \sqrt{2 + \gamma_0}$ inside the band gap. (c) Profile of the topological state localized at the left end of the chain for $\gamma=0.4$. Inset shows the zoomed-in view of the edge state corresponding to the shaded area. Due to the chiral symmetry, the state has a zero displacement at its even sites.}
\label{fig1}
\end{figure} 

Our system consists of a chain of particles interconnected with two alternating linear springs and grounded with nonlinear springs as shown in Fig.~\ref{fig1}a.
The stiffnesses of the two linear springs are $1-\gamma$ and $1+\gamma$ in normalized units. The ground springs are characterized by a linear stiffness of $\gamma_0$ and a parameter $\Gamma$  that introduces cubic nonlinearity. The dynamics is thus governed by the following second-order ODEs in time:

\begin{equation}
\left.
\begin{IEEEeqnarraybox}[
\IEEEeqnarraystrutmode
\IEEEeqnarraystrutsizeadd{2pt}
{2pt}
][c]{rCl}
 \ddot{x}_{j} &=& (1+\gamma) (x_{j-1}-x_{j}) - (1-\gamma) (x_{j}-x_{j+1}) \\
&& - \gamma_0 x_{i} -\Gamma x_{i}^3, \\ 
 \ddot{x}_{j+1} &=&(1-\gamma) (x_{j}-x_{j+1})  - (1+\gamma) (x_{j+1}-x_{j+2}) \\
 && - \gamma_0 x_{j+1} - \Gamma x_{j+1}^3,
\end{IEEEeqnarraybox}
\, \right\}
\label{eq:eom_KG}
\end{equation}

\noindent where $x_{j}$ and $x_{j+1}$ denote the dynamic displacements of two masses inside the unit cell; overdots represent the derivatives with respect to time $t$. 

The linear component of the system ($\Gamma=0$) is associated with the following eigenvalue problem:
\begin{equation}
\omega^2 \mathbf{X} =  \mathbb{D} \mathbf{X}, 
\end{equation}

\noindent where $\omega$ and $\mathbf{X}$ denote the eigenfrequency and eigenvector of lattice vibrations, respectively, and the dynamical matrix
\begin{equation}
\mathbb{D} = 
\begin{bmatrix}
2 + \gamma_0 & -(1 - \gamma) & ... & 0 & 0 \\
-(1 - \gamma)  &  2+ \gamma_0 & -(1 + \gamma) &...  & 0 \\
... & ... & ... & ... & ... \\
0 & ... &-(1 + \gamma) &  2+ \gamma_0 & -(1 - \gamma) \\
0 & 0 & ... & -(1 -		 \gamma) & 2+ \gamma_0
\end{bmatrix},
\end{equation}

\noindent for a chain that is fixed on its left and right ends and consists of $n \in 2 \mathbb{Z}$ particles. After the removal of the diagonal of this dynamical matrix, which simply shifts the spectrum to a non-zero (finite) frequency, i.e., $\omega^2 = 2 + \gamma_0$, the chiral symmetry of the remaining matrix can be observed \cite{Prodan2009, Susstrunk2016}. In other words, we have an anti-commutative relation 
\begin{eqnarray}
\mathrm{\Sigma}_z  \left [ \mathbb{D} - (2 + \gamma_0)    \mathbb{I} \right]  +  \left [ \mathbb{D} - (2 + \gamma_0) \mathbb{I} \right ] \mathrm{\Sigma}_z =0,
\label{eq:anticomm}
\end{eqnarray}

\noindent where $\mathbb{I} $ denotes the identity matrix and $ \mathrm{\Sigma}_z$ is constructed from the internal chiral operator $\sigma_z = \begin{bmatrix}
1 & 0 \\
0 & -1
\end{bmatrix} $ 
for the unit cell, such that $  \mathrm{\Sigma}_z= \sigma_z \oplus \sigma_z \oplus ... \oplus \sigma_z$. 

As a result of the chiral symmetry, the spectrum ($\omega^2$) is symmetric around  the mid-gap frequency ${\omega^2 = 2 + \gamma_0}$.
For a nonzero $\gamma$, the system supports a band gap centered at the mid-gap frequency, which, in our case, is independent of $\gamma$. 
We take a large chain of particles ($n=100$) with fixed and free boundary conditions on the left and right ends, respectively, and plot its spectrum in Fig.~\ref{fig1}b. Note that we keep the right end free, which slightly breaks the chiral symmetry, to focus on the states only on the left end as discussed next. As we increase $\gamma$, we observe that the band gap closes and opens again, leading to the so-called band inversion \cite{Chaunsali2017}. The system also makes a topological transition at $\gamma=0$ that is quantified by the change in the topological invariant calculated from the bulk dispersion properties. The physical implications of this are reflected in the spectrum of the finite chain when we observe a state emerging inside the band gap for $\gamma>0$ at the mid-gap frequency. 
This state is localized at the left boundary of the chain (see Fig.~\ref{fig1}c) and emerges due to the nontrivial topology of the bulk for $\gamma>0$. Moreover, chiral symmetry of the dynamical matrix imparts a special profile to the edge state. It has zero displacement at even sites as shown in the inset. 

In what follows we will focus on this edge state and study the effect of nonlinearity on its frequency, shape, and stability. We fix $\gamma=0.4$ (marked in Fig.~\ref{fig1}b) to ensure a large band gap in the linearized spectrum. This accommodates the effect of strong nonlinearity by which the frequency of the edge state can vary considerably within the band gap. Also, the effects of the right edge on its dynamics are negligible since we consider a long chain and the linear edge state shown in Fig.~\ref{fig1}c is well localized on the left. We choose the linearized ground stiffness $\gamma_0 = 1$ for ease, but also discuss the implications when we tune this parameter.

 \begin{figure*}[!]
\centering
\includegraphics[width=\textwidth]{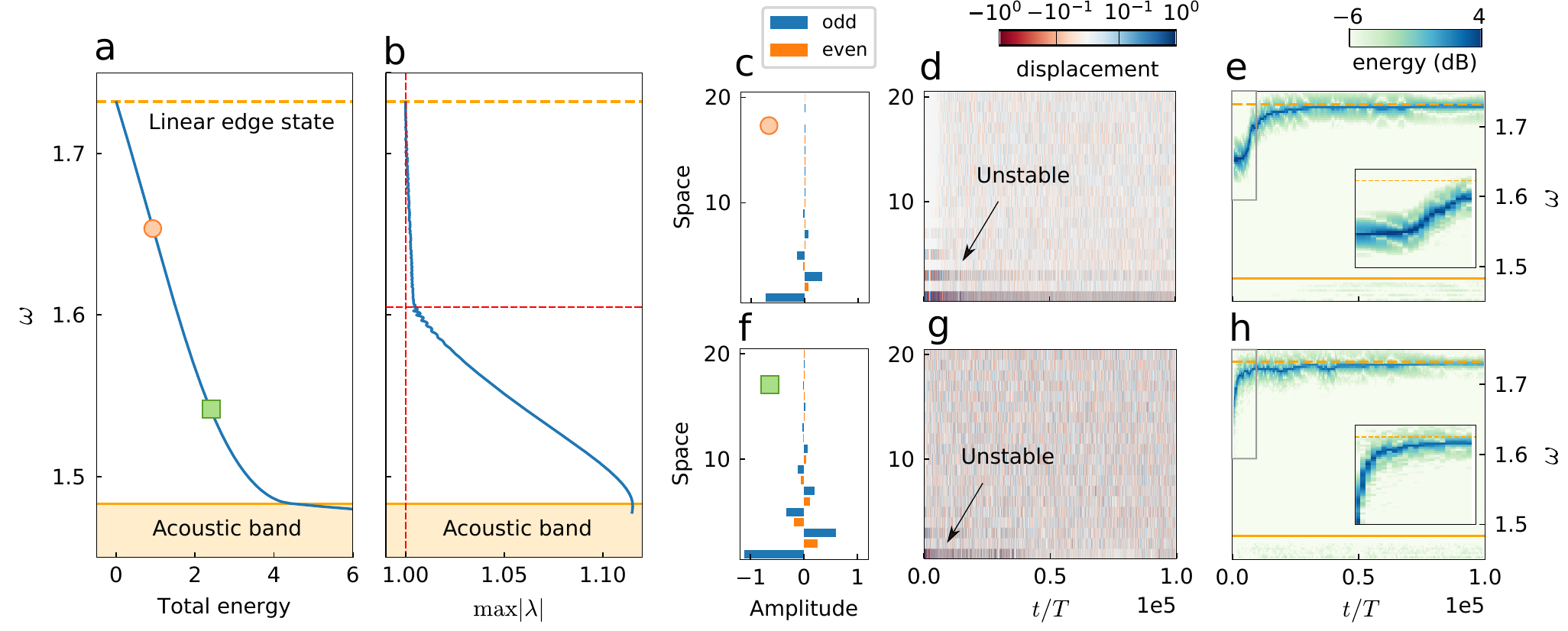}
\caption{Effect of \textit{softening} nonlinearity (with $\gamma_0=1$, $\gamma=0.4$ and $\Gamma=-0.8$). (a) Nonlinear continuation of the topological edge state. (b) Maximum amplitude of FMs of the periodic solutions deviating from unity (vertical dashed line) with the decrease in  frequency. The horizontal dashed line in red marks the onset of high growth rate of instability. (c) Profile of the nonlinear state at $\omega=1.65$ as indicated by the circular marker in (a). Its even sites have a non-zero displacement. (d) Transient simulation with the nonlinear state as the initial condition. (e) STFT of the first particle to indicate the change in frequency over a long time. The inset shows the zoomed-in view corresponding to the box. (f)--(h) Same for the nonlinear state at $\omega=1.54$, which has a stronger instability.}
\label{fig2}
\end{figure*} 

 \begin{figure*}[!]
\centering
\includegraphics[width=\textwidth]{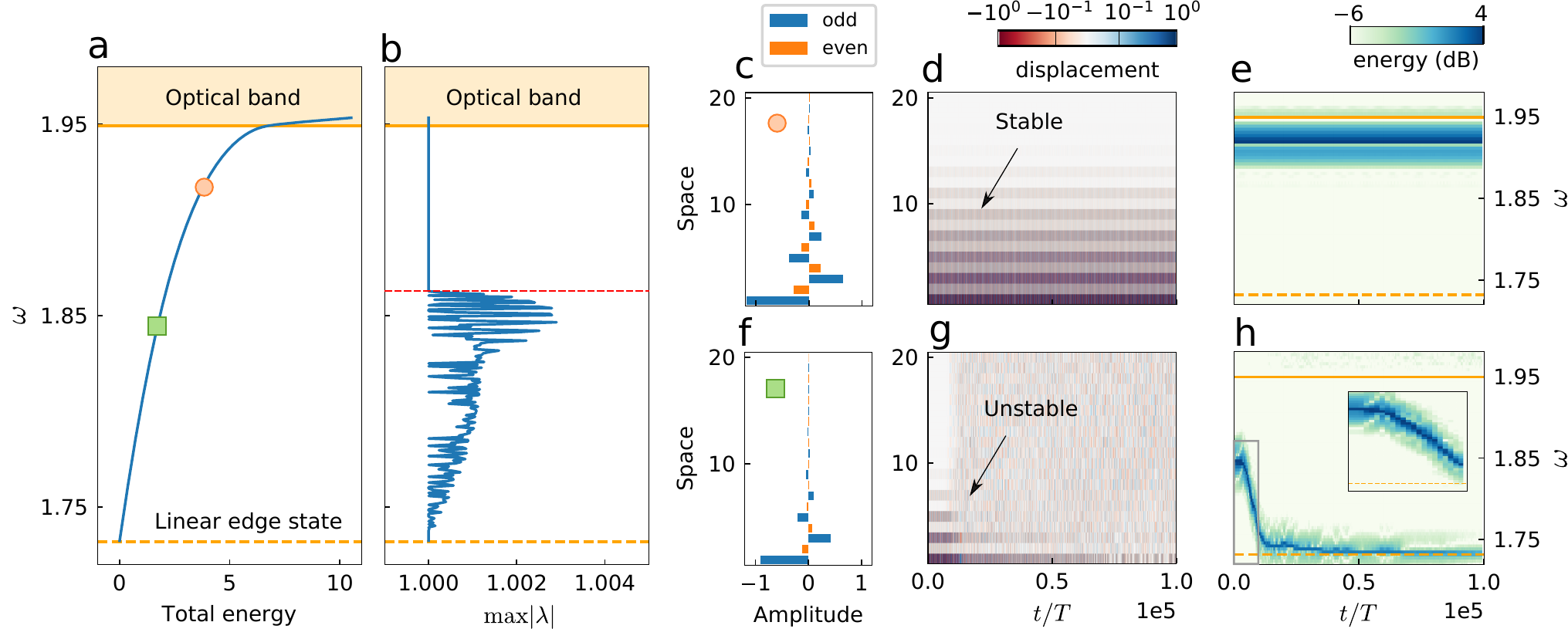}
\caption{Effect of \textit{stiffening} nonlinearity (with $\gamma_0=1$, $\gamma=0.4$ and $\Gamma=0.8$). (a) Nonlinear continuation of the topological edge state. (b) Maximum amplitude of FMs of the time-periodic solutions. Note the region above the horizontal dashed line in red supports large-amplitude nonlinear states that are linearly stable. (c) Profile of the nonlinear state at $\omega=1.92$ as indicated by the circular marker in (a). (d) Transient simulation with the nonlinear state as the initial condition. Due to linear stability, the state remains localized for a long time without shedding its energy into the bulk. (e) STFT of the first particle verifies that there is no energy transfer across different frequencies. (f)--(h) Same for the nonlinear state at $\omega=1.85$; however, in this case the instability of the state causes the delocalization and change in frequency of the edge state over a long time.} 
\label{fig3}
\end{figure*} 

\section{Nonlinear continuation and dynamics} 
We take the linear edge state as the initial condition in the nonlinear Newton solver to find the family of nonlinear periodic solutions at a frequency that is varied in small steps. We first consider \textit{softening} nonlinearity ($\Gamma < 0$). In Fig.~\ref{fig2}a, we show the decrease in the edge-state frequency as the total energy of the system increases.  Moreover, for large energy the state penetrates the acoustic band below and resonates with bulk linear states as reflected by the sudden rise in energy inside the band. This has been observed in a recent experiment \cite{Vila2019}. However, even before the state penetrates into the bulk spectrum, it can develop various instabilities giving rise to rich dynamics. To examine the linear stability of these periodic solutions, we use the Floquet theory (see Appendix \ref{Appendix_A}). 
Recall that FMs of modulus larger than unity imply exponential instability, whereas those with a unit amplitude imply linear stability, which guarantees a long life time of the nonlinear state~\cite{Aubry2006}. 
In Fig.~\ref{fig2}b, we show the maximum amplitude of FMs, i.e., max($|\lambda|$) (in total we have $2n$ FMs), as a function of frequency. We know that the linear edge state with $\omega_{ed}= \sqrt{2+\gamma_0} = 1.73$ must have all its FMs with unit amplitude. However, its nonlinear continuation reveals that max($|\lambda|$) deviates from unity as the frequency decreases, indicating the presence of instabilities. At about $\omega=1.6$, we notice a sudden change in the growth rate of instability, which is linked to the onset of a different type of instability. This will be further discussed in the following section.

Next, we examine how the aforementioned instabilities affect the dynamics of the nonlinear edge state. In Fig.~\ref{fig2}c, we show the shape of the nonlinear edge state at $\omega=1.65$ with max$\abs{\lambda}=1.002$. One can see that nonlinearity modifies the shape of the edge state. In particular, the even sites now obtain a non-zero displacement. We use this as the initial condition of our system, then add white noise with $1 \%$ amplitude of displacement, and perform  Runge-Kutta simulations for a long time of $10^5 T$, where $T$ denotes the time period of the periodic solution. Due to instability, this nonlinear edge state soon disperses its energy into the bulk as shown in Fig.~\ref{fig2}d. Consequently, the edge state lowers its energy and tends toward the shape and frequency of the \textit{linear} edge state. This fact is demonstrated more clearly by performing the Short-Time Fourier Transformation (STFT) on the displacement of the first particle (with the time window of $500 T$). We see the up-shift of frequency to the linear edge-state frequency over a long time in Fig.~\ref{fig2}e. The inset, which represents the dynamics for the time of $10^4 T$, shows that the instability takes about $5000 T$ to manifest and shift the frequency.

To study the case of stronger instabilities, we now take a nonlinear edge state (at $\omega=1.54$) with larger FM (max($\abs{\lambda}=1.062$)), shown in Fig.~\ref{fig2}f. Again, looking the profile of this nonlinear edge state, we note non-zero displacements at even sites. We use this state as an initial condition by adding white noise with 1\% amplitude of the displacement.
Similarly to the previous case, this state quickly disperses its energy into the bulk (Fig.~\ref{fig2}g). Its frequency again springs back to the initial linear edge-state frequency, which is shown in Fig.~\ref{fig2}h; however, this transition is faster compared to the one shown in Figs.~\ref{fig2}d and \ref{fig2}e because of the stronger instability. We roughly estimate that it takes $160 T$ to manifest the same relative growth in the initial state [see Eq.~\eqref{eq:growth} in Appendix \ref{Appendix_A}] as in the previous case at $\omega=1.65$. This makes sense as the inset of Fig.~\ref{fig2}h shows the up-shift of frequency right from the first window of STFT, which is $500 T$ long.

It is worth highlighting that when the nonlinear edge state delocalizes by shedding its energy and tends to the linear edge state in a long time, a part of its energy transfers to all the bulk modes in the system.
It remains to be studied in the future if such a scenario leads to a thermal equilibrium and whether the effect of strong nonlinear interactions leads to an effective renormalization of the linear dispersion relation \cite{Gershgorin2005, Jiang2014}. 

Now we study the case of \textit{stiffening} nonlinearity, i.e., with $\Gamma > 0$. In Fig.~\ref{fig3}, we show the nonlinear continuation and dynamics of the linear edge state for this case. With the increase in energy, the frequency of the nonlinear edge state now increases (Fig.~\ref{fig3}a). However, upon examining its stability, we observe that the maximum amplitude of FMs shows a remarkably different trend from the \textit{softening} case, shown in Fig.~\ref{fig2}b. We find that not only is the instability smaller by one order of magnitude in the associated FMs (when 
the instability is present), but also there exists an  interval of no instability (i.e., $|\lambda| =1$) for frequencies more than about $\omega=1.86$. Therefore, we discover a region where \textit{high-amplitude edge states can be linearly stable} for stiffening nonlinearity within our dimer chain. 
We show one such nonlinear state at $\omega = 1.92$ in Fig.~\ref{fig3}c. Once again, we observe that the even sites develop a non-zero displacement. 
We demonstrate its linear stability by performing a long-time simulation shown in Figs.~\ref{fig3}d, and \ref{fig3}e. Evidently, this nonlinear edge state remains localized on the boundary without shedding any energy into the bulk. On the other hand, a \textit{low-amplitude}  edge state (Fig.~\ref{fig3}f) at $\omega=1.85$ lacks this stability and disperses its energy into the bulk (Fig.~\ref{fig3}g) with its frequency springing back to the initial linear-state frequency (Fig.~\ref{fig3}h).

To sum up, in general, nonlinearity leads to deformation in the shape of topological edge states, e.g., emergence of non-zero displacements at its even sites in this case, and shifts its frequency. Most importantly, nonlinearity also leads to instabilities. Here, we showed that the presence of onsite nonlinearity of a cubic form leads to instabilities, which disperse the energy from the edge to the bulk of the lattice  over long times. However, we also found a frequency regime, with strongly nonlinear dynamics, where nonlinear edge states are linearly (and dynamically in our
direct numerical 
simulations) stable. This provides the opportunity to trap significant amounts of energy at the boundary for long time intervals. We also observed that stability trends are drastically different when we use different types of nonlinearity, i.e., softening or stiffening, even though we consider a simple cubic functional form. 
This suggests that the effect of nonlinearity on topologically-nontrivial
systems strongly depends on the details (type, form, and location) of the 
nonlinearity. In the following section, this will be more evident when we explore several different types of instabilities emanating from the complex interaction between extended and local modes in the phase diagram of FMs.

\begin{figure}[!]
\centering
\includegraphics[width=\columnwidth]{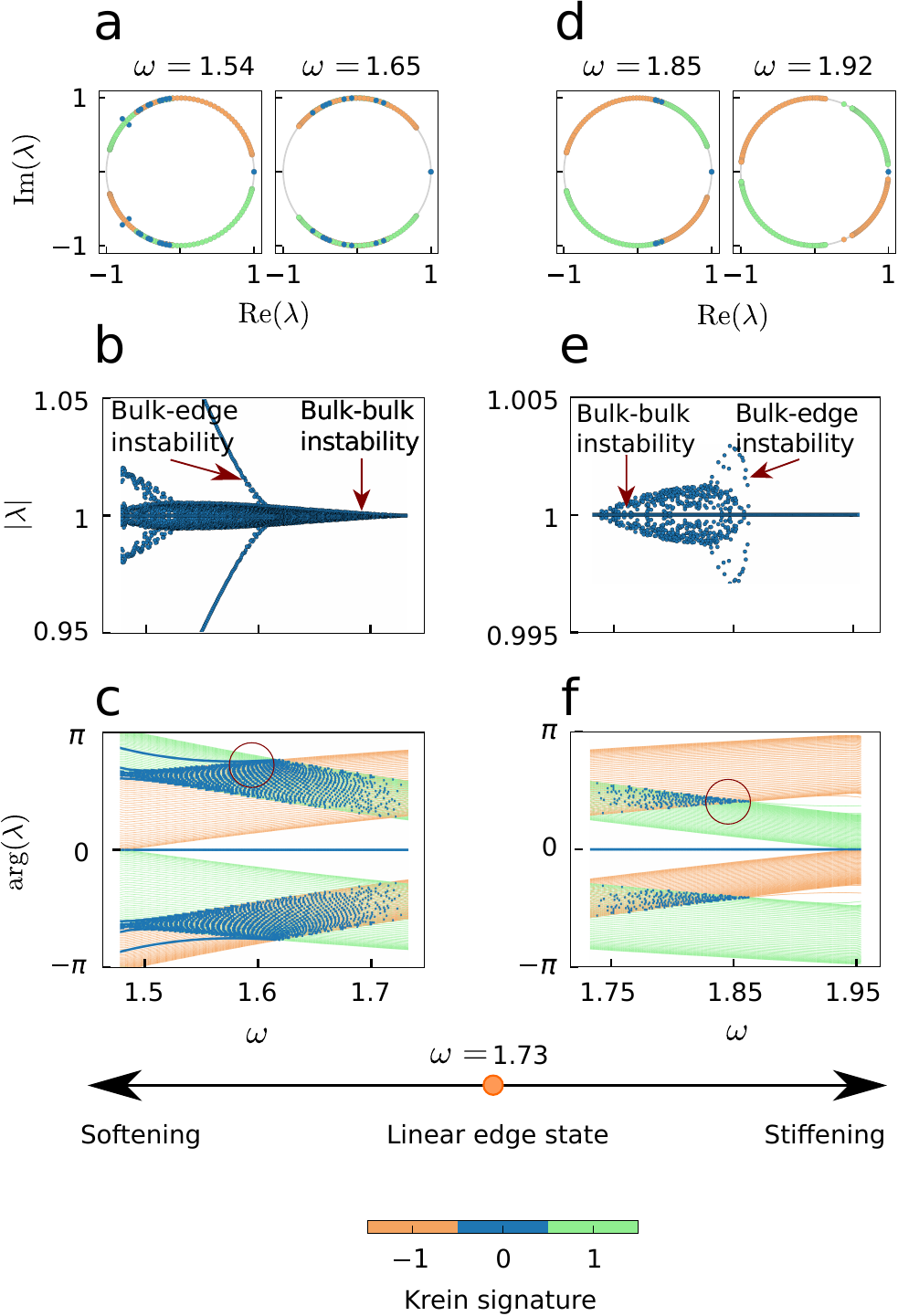}
\caption{Variation of FMs with frequency. 
(a) FMs in the complex plane for the two cases discussed in Fig.~\ref{fig2} for \textit{softening} nonlinearity. Color denotes the Krein signature of each FM.
(b) Variation in the amplitude of FMs with frequency. Values away from $|\lambda|=1$ indicate the presence of instability.  
(c) Variation in the phase of FMs with frequency. Blue dots denote the unstable FMs with zero Krein signature. The circled region represents the onset of a different instability.
(d)--(f) The same for \textit{stiffening} nonlinearity.}
\label{fig4}
\end{figure} 

\section{Linear stability}
In this section, we investigate the linear stability of the nonlinear edge state in more detail especially to answer the following questions: What \textit{type} of instabilities exist in the system and do they differ in case of \textit{softening} and \textit{stiffening} nonlinearities? Since there are $2n$ complex FMs for each periodic solution, we plot them all in Fig.~\ref{fig4}.
First, for the \textit{softening} case, we plot all FMs in the complex plane in Fig.~\ref{fig4}a for $\omega=1.54$ and $\omega=1.65$, the two cases with instability  previously discussed in Fig.~\ref{fig2}. We also plot a unit circle to guide the eye.
In a linearly stable system, all the FMs must lie on the unit circle for a Hamiltonian system as ours. Since FMs are the eigenvalues of the monodromy matrix (Appendix \ref{Appendix_A}), which is real, these come in complex conjugate pairs. Note that there is a pair of FMs always located at the point ($+1,0$) of the unit circle. These correspond to the phase mode \cite{Aubry2006}. 
Instability is caused as some FMs leave the circle. There are two distinct ways for that to happen. The first way is when two complex conjugate  FMs collide on the real axis, i.e., either at ($+1,0$) or ($-1,0$),  and leave the unit circle to remain on the real axis. These are often termed as ``real'' instabilities and are typically independent of the size of the system. The second way is when FMs collide \textit{elsewhere} on the unit circle. In this case, two pairs of complex conjugate FMs collide in such a way that a quadruplet of FMs leave the circle. Due to the symplectic property of our Hamiltonian system, these four FMs come in conjugate and reciprocal pairs, i.e., $\lambda$, $1/\lambda$, $\lambda^*$, and $1/\lambda^*$, where $*$ denotes complex conjugation. This scenario causes ``oscillatory'' or Krein instabilities.

A necessary condition for 
an instability to occur is that the colliding FMs must have opposite Krein signatures \cite{Marin1998, Aubry2006, Flach2008}. 
Krein signature corresponding to a FM ($\lambda$) can be calculated from its eigenvector ($v$) as

\begin{eqnarray}
K(\lambda)=\mathrm{sgn}\left[ v^{\dagger} (-i J) v \right],
\end{eqnarray}
\noindent  where $i$ and $\dagger$ denote imaginary unity and  complex transpose, respectively; and 
$ J = 
\begin{bmatrix}
0 & \mathbb{I}_n \\
-\mathbb{I}_n & 0
\end{bmatrix}
$ with $\mathbb{I}_n$ being the unit matrix of dimension $n$. Physically, the Krein signature represents the sign of energy associated with the eigenvector \cite{Chung2020}. It is either $+1$ or $-1$ for the FMs lying on the unit circle, except those on the real axis, for which it is 0 by definition. Moreover, the FMs leaving the unit circle also have vanishing Krein signature and we will use this fact to identify instabilities in the discussion below.

Going back to Fig.~\ref{fig4}a,  we color each FM as per its Krein signature. The FMs with opposite Krein signatures (orange and green arcs) collide and leave the unit circle to cause instabilities. The unstable FMs are shown in blue except for the phase mode at (+1,0). We clearly observe that in both cases, for $\omega=1.54$ and $\omega=1.65$, the system displays oscillatory instabilities. For $\omega=1.54$, the instability is larger, as also observed in the previous section, and therefore, the FMs leaving the unit circle are more clearly discernible. 

To investigate if these cases differ in any other way, we keep track of all FMs by sweeping the frequency of the nonlinear edge state. In Figs.~\ref{fig4}b and \ref{fig4}c, we plot the amplitude and phase of FMs at various frequencies of the nonlinear edge state. As the frequency decreases from $\omega_{ed} = 1.73$ of the linear edge state,  we observe $|\lambda|$ deviating from unity (or $\lambda$ moving away from the unit circle in the complex plane) and manifesting as instabilities of different types. We further verify this by looking at arg($\lambda$) variation, in which, for $\omega \gtrapprox 1.6$, the spectral bands of opposite Krein signatures (in green and orange) collide and cause instabilities (in blue). We call them ``bulk-bulk'' instabilities since these emerge due to the collision between two extended (bulk) states. Since the number of extended states depends on the size of the system, this instability occurs more often in frequency as the length of chain increases but decreases in its strength (not shown here). Indeed, we expect these instabilities to disappear in the infinite lattice limit. Therefore, these are often termed as \textit{finite-size} instabilities \cite{Marin1998}. Note that the spectral bands manifest as the arcs on the unit circle shown in Fig.~\ref{fig4}a at a given frequency. These are also related to acoustic and optical branches of dispersion of the linearized system (see Appendix \ref{Appendix_B} for more details).

However, for $\omega \lessapprox 1.6$, the dominant instability stems from the collision of an isolated FM, which bifurcates from one spectral band (circled area in Fig.~\ref{fig4}c), with the other band of opposite Krein signature, and both escaping the unit circle. The eigenvectors of the bifurcated FMs are spatially localized, and these are often referred as ``internal modes'' of the system \cite{Aubry2006}. We call the resulting instability as ``bulk-edge'' instability since it is caused by the collision between an extended (bulk) and a localized (edge) state. Onset of this instability thus explains the sudden rise in instability growth observed in Fig.~\ref{fig2}b earlier and the distinct instability strengths for the two cases at $\omega=1.54$ and $\omega=1.65$.

Now we follow a similar line of investigation for the system with \textit{stiffening} nonlinearity and show the results in Figs.~\ref{fig4}d--f. 
Clearly, the case with frequency $\omega=1.85$, which showed instability in Fig.~\ref{fig3}, possesses an oscillatory type of instability because collisions of FMs do not occur on the real axis in Fig.~\ref{fig4}d. However, the case with frequency $\omega=1.92$ is linearly stable since FMs do not collide on the unit circle. To further explain these scenarios, we look at the variation of FMs with frequency. 
In Figs.~\ref{fig4}e--f, as the frequency increases from $\omega_{ed} = 1.73$ of the linear edge state, we first notice the ``bulk-bulk'' instability that is caused by the collision of extended bulk states of two spectral bands. At $\omega \approx 1.85$ we notice an onset of a different instability. In the circled area of Fig.~\ref{fig4}f, a state bifurcates from the upper limit of the spectral band (green) and collides with the extended states of the other band (orange). We thus call this as ``bulk-edge'' instability. For frequencies beyond about $\omega = 1.86$, we clearly notice that there are no instabilities in the system. Neither the spectral bands intersect to cause the ``bulk-bulk'' instability nor the bifurcated modes collide with the bands to cause any ``bulk-edge'' instability. We have verified that this range remains intact even in the presence of large chain of 500 particles. This, therefore, explains our earlier observations in Fig.~\ref{fig3}b that the nonlinear state at $\omega=1.85$ is unstable whereas the state at $\omega=1.92$ is linearly stable.

In this section, we have thus seen different types of instabilities that manifest with the change of nonlinearity. Those can be systematically analyzed with the phase and Krein signature of the FMs, as these offer insight on the nature (and thresholds) of the instabilities. 
With these tools at hand, one can further change system parameters, such as $\gamma$, $\gamma_0$, and the functional form of nonlinearity, to achieve a desired property of the nonlinear edge state. For example, one could, in principle, tailor the parametric stability intervals by shifting them to different frequency ranges. Moreover, the functional form of the nonlinearity can be changed to control the bifurcation of isolated FMs that cause instabilities. Though most of these tasks primarily rely on numerical techniques due to the strongly nonlinear nature of the system, one can still employ analytics to gain insights into some of these phenomena. For example, as stated earlier, the spectral bands in the phase diagram of FMs (Figs.~\ref{fig4}c,f) are related to the dispersion properties of the linearized system (around the uniform
vanishing displacement state). Therefore, one can get an analytical estimate of their variation with frequency and predict the regions where those do not overlap to cause ``bulk-bulk'' instabilities. For the stiffening case, our analytical calculation predicts the region of stability for $\omega > (\omega_1+\omega_c)/2 = 1.86$, where $\omega_1$ and $\omega_c$ represent the upper cutoff frequencies of acoustic and optical bands, respectively (see Appendix \ref{Appendix_B} for details). This closely matches with our numerical results discussed in Fig.~\ref{fig4}f.

 \begin{figure}[!]
\centering
\includegraphics[width=\columnwidth]{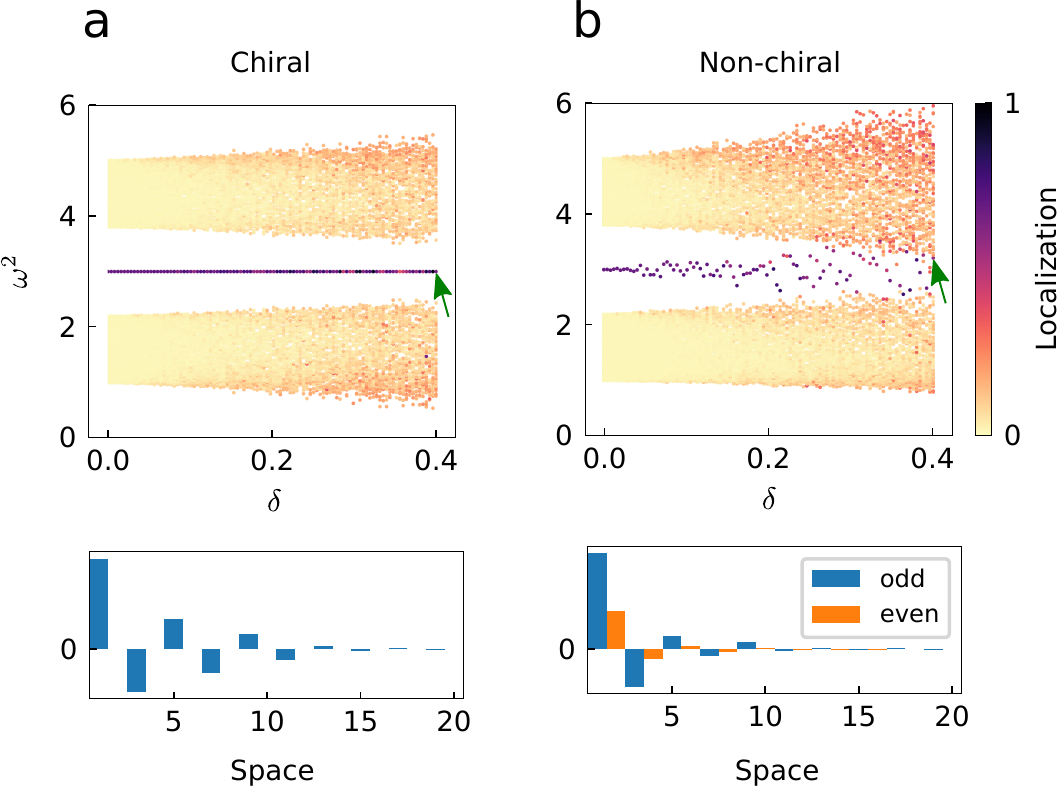}
\caption{Effect of disorder on the spectrum of the linearized system. (a) Chiral disorder preserves the frequency of the edge state ($\omega^2_{ed}=3$) inside the band gap. Also, the bands above and below remain symmetric about the edge-state frequency. Colormap indicates the localization (IPR) of states. Below is a typical shape of the edge state where even sites have zero displacements at $\delta=0.4$. (b) Non-chiral disorder does not impose any such symmetry constraints on the spectrum and the frequency of the edge state changes with disorder. Moreover, the edge state develops a non-zero displacement at even sites in the presence of disorder ($\delta=0.4$).}
\label{fig5}
\end{figure} 

\begin{figure*}[!]
\centering
\includegraphics[width=\textwidth]{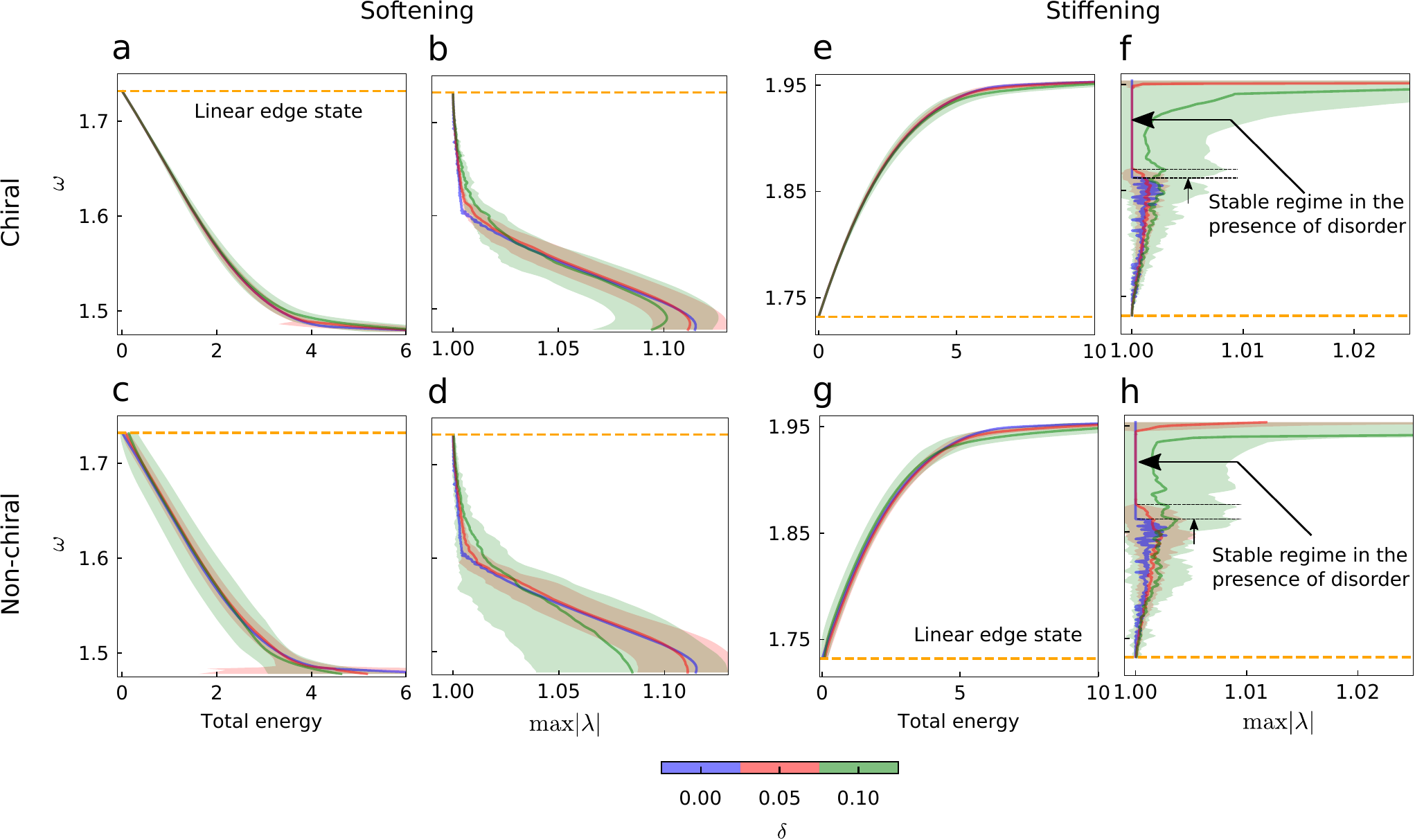}
\caption{Effect of disorder on the nonlinear continuation and instabilities of the edge state. (a)--(d) For the \textit{softening} case when the system has chiral (a, b) and non-chiral (c, d) disorder. Colors indicate disorder strengths of 0 (pristine), 0.05, and 0.1. Standard deviations in $x$-axes are shown by the shaded areas. (e)--(h) The same for the \textit{stiffening} case. Notice the region of stability for $5\%$ disorder. A slight reduction in its  frequency span compared to the pristine case is marked with horizontal dashed lines.}
\label{fig6}
\end{figure*} 

\section{Effect of disorder}
In the previous sections, we have discussed the effect of nonlinearity on the edge-state frequency and stability. In this section, we ask how these trends vary if we introduce a weak disorder into the chain. This question is relevant since topological edge states are known to be robust against certain types of disorder in linear regimes. However, it is not clear if such arguments could be generalized in nonlinear regimes as we discuss here. Generally speaking, the presence of disorder would alter the linear spectrum of the system. This includes the change in cutoff frequencies and spatial localization of eigenvectors known as Anderson localization. We, therefore, expect that the presence of disorder will be reflected in the variation of linear stability of the nonlinear edge state (We saw how spectral bands and the bifurcations of localized FMs from them dictate stability in Figs.~\ref{fig4}b and \ref{fig4}d). As we show below, introducing different \textit{types} of disorder into the chain will help us understand the problem in more detail.

We discussed earlier in this manuscript, that the dynamical matrix $\mathbb{D}$ of the linearized system obeys chiral symmetry after we remove its diagonal. Consequently, a disorder that respects chiral symmetry, namely ``chiral disorder'', will alter the spectrum in such a way that the edge state remains fixed to the mid-gap frequency $\omega_{ed}$ and localized at the edge, and hence will be robust. Moreover, the entire spectrum ($\omega^2$) will remain symmetric about this frequency. To introduce chiral disorder, we need to perturb the stiffnesses of springs in a manner such that the diagonal elements of the resulting dynamical matrix $\mathbb{D}_{\delta}$ remain \textit{independent} of the perturbation, and again satisfy Eq.~\eqref{eq:anticomm}. This is where we appreciate the need of having ground springs in the system so that chiral disorder can be introduced and systematically studied. 
More precisely, in Fig.~\ref{fig1}a, if the stiffness of red and blue springs are perturbed as $(1+\gamma) \rightarrow (1+\gamma + \delta_1)$ and $(1-\gamma) \rightarrow (1-\gamma + \delta_2)$, we change the stiffness of ground spring (green) as $\gamma_0 \rightarrow (\gamma_0 - \delta_1 - \delta_2)$. We use the disorder magnitude $\delta$ to randomize $\delta_1$ and $\delta_2$, such that  $\delta_{1,2} = \delta \times \text{rand}(-1,1)$. 
As a result, the diagonal of the resulting dynamical matrix $\mathbb{D}_{\delta}$ does not change. 
It is evident that this type of disorder is specialized and hard to achieve experimentally, yet still interesting theoretically.

We demonstrate the effect of this disorder in Fig.~\ref{fig5}a by plotting eigenfrequencies and localization of corresponding eigenvectors as the disorder magnitude $\delta$ is increased. For an eigenvector with $n$ masses, we use the localization index (i.e., inverse participation ratio) defined as 
\begin{eqnarray}
\text{IPR}=\frac{\sum\limits_{j=1}^{n}{u_j^4}}{\left( \sum\limits_{j=1}^{N}{u_j^2} \right )^2}, 
\end{eqnarray}
\noindent where $u_j$ denotes the displacement of the $j$th mass. 
Evidently, the edge state inside the band gap stays at $\omega_{ed}^2 = 3$ for any amount of disorder $\delta$, hence it is robust. It is localized on the edge of the system and maintains its typical shape (zero displacement at even sites) governed by the chiral symmetry.  Also, the band spectrum ($\omega^2$), i.e., acoustic and optical bands, remains symmetric about $\omega_{ed}^2$, irrespective of the magnitude of disorder (a slight breaking of symmetry is due to the free boundary condition on the right end of the chain). 
In Fig.~\ref{fig5}b, we show variation of the spectrum when disorder is introduced in all the springs randomly. We call this a ``non-chiral'' disorder. We observe that the edge state inside the band gap no longer remains fixed at one frequency and exhibits non-zero displacements at its even sites. Moreover, the bands lose the symmetry as we increase disorder strength.

In both of the cases above, we notice a change in frequency cutoffs and an increase in spatial localization of eigenvectors. These factors will be important in explaining the nonlinear continuation of the edge state discussed below.  In particular, we monitor how $\omega_1$ and $\omega_c$, the upper cutoff frequencies of acoustic and optical bands, respectively, vary with the increase in disorder strength to estimate the region with no ``bulk-bulk'' instability in the \textit{stiffening} case. We perform 50 disorder realizations for every strength of disorder $\delta \in [0, 0.1]$ to get a second-degree polynomial fit for the \textit{maximum} of functions $\omega_1$ and $\omega_c$. We choose the maximum since we want to get the most conservative estimate of the stability region, i.e., $\omega > (\omega_1+\omega_c)/2$. For chiral disorder, we get
\begin{eqnarray}
\omega_1(\delta) &=&  0.87 \delta^2 + 0.17\delta + 1.48, \label{eq:cw1}\\
\omega_c(\delta) &=& 0.68 \delta^2 + 0.14 \delta +  2.23, \label{eq:cwc}
\end{eqnarray}

\noindent and for non-chiral disorder:
\begin{eqnarray}
\omega_1(\delta) &=& 0.46\delta^2 + 0.34 \delta +  1.48, \label{eq:rw1} \\
\omega_c(\delta) &=& 1.61 \delta^2 + 0.37\delta + 2.23.\label{eq:rwc}
\end{eqnarray}

We now introduce the aforementioned types of disorder into the chain and perform the nonlinear continuation of the edge state at three different disorder magnitudes: $\delta= 0$, 0.05, and 0.1. These are normalized to unity -- the mean stiffness of our dimer system.  In Fig. \ref{fig6}a, we plot the mean and standard deviation of the total energy of nonlinear edge states in the case of \textit{softening} nonlinearity and chiral disorder. 
Due to the softening effect, and as we showed before for the defect-free case in Fig. \ref{fig2}a, the frequency of the nonlinear state decreases as the total energy increases. When disorder is increased, we notice that the mean of total energy remains relatively close but standard deviation widens as we come down on the y-axis. In Fig. \ref{fig6}b, we observe similar trends for the mean and standard deviation of maximum FMs against frequency. Note that the standard deviation approaches to zero for the linear edge-state frequency $\omega_{ed}$ in these plots. This is the direct consequence of having a chiral disorder, which does not change the linear edge-state frequency for any amount of disorder strength.  

In Figs. \ref{fig6}c,d, we show the trends for non-chiral disorder. 
Here we have relied on the interpolation of the nonlinear continuation curves in the frequency steps that exactly match with the earlier case in Figs. \ref{fig6}a,b.
This is because a non-chiral disorder does not impose any symmetry restrictions on the linear spectrum; it changes the linear edge-state frequency, which is the starting point for the nonlinear continuation. 
We observe that the deviations are generally larger than those seen for the case of chiral disorder, implying that the nonlinear continuation is less robust for non-chiral disorder than it is for chiral disorder.

We perform similar calculations for the \textit{stiffening} case and plot those in Figs. \ref{fig6}e--h. We reach the same conclusion that the nonlinear continuation and stability are more robust (with lesser fluctuations) in the presence of chiral disorder.
However, the upper region where high-amplitude edge states were found to be linearly stable earlier shows an interesting trend.
The region of stability remains almost intact (i.e., the mean remains unity with a zero standard deviation) even in the presence of 5\% disorder of both types. Yet, there is a slight reduction in its frequency span, which is more for non-chiral disorder ($\Delta\omega = 0.015$) than chiral disorder ($\Delta \omega = 0.008$). These are the up-shifts of dashed lines shown in Figs. \ref{fig6}f,h. 

To understand what types of instabilities manifest in this region, we look back at Fig.~\ref{fig4}f. For $\omega>1.86$, the spectral bands do not overlap. Also, the bifurcated modes do not interact among themselves or collide with any of the spectral bands. We know that the presence of disorder is reflected in two ways. The linearized spectrum shows a deviation in its cutoff frequencies and the eigenvectors tend to be more localized. The former leads to a gradual \textit{decrease} in the stability region by enhancing ``bulk-bulk'' instability. In other words, the intersection of bands in the circled area occurs due to the spreading of band spectrum. Moreover, the latter affects the bifurcation and collision associated with the localized FM. For 5\% disorder, we observe that the former scenario is responsible for enhancement in the range of ``bulk-bulk'' instabilities. Since we have already calculated the shifts in cutoff frequencies in Eqs.~\eqref{eq:cw1}--\eqref{eq:rwc}, we provide a conservative estimate of reduction in stability region as

\begin{eqnarray}
\Delta \omega(\delta) &=& \omega(\delta) - \omega(0)  \nonumber \\
&=& \left[ \frac{\omega_1(\delta) + \omega_c(\delta)}{2} \right]-  \left[ \frac{\omega_1(0) + \omega_c(0)}{2} \right],
\end{eqnarray} 

\noindent which is 0.01 and 0.02 for chiral and non-chiral types of disorder, respectively, at $\delta=0.05$. This explains our observation of a larger reduction in the stability range in the presence of non-chiral disorder.
  
For 10\% disorder, the region in Figs. \ref{fig6}f,h displays instability. We have verified that all the instabilities are related to the  bifurcation and collision of the localized FM. This, therefore, indicates the dominant role of localized eigenvectors in the presence of disorder. In this regime of disorder, we do not observe any specific difference between the cases with chiral or non-chiral disorder. 

We, therefore, conjecture that with an increase in the disorder strength the range of linear stability of the nonlinear edge state \textit{gradually} reduces (see Appendix \ref{Appendix_C} for further support). This reduction is due to the gradual shift in cutoff frequencies in the presence of disorder. This enhances the frequency range of ``bulk-bulk'' instabilities. However, there is a threshold of disorder after which the localization of eigenvectors plays an important role and instabilities are caused by the change in the bifurcation and collision patterns of the localized FMs, which need not be monotonous with disorder strength.

\section*{Conclusions}
We consider a one-dimensional SSH-like chain that hosts a topological edge state at finite-frequency and study the effect of strong nonlinearity on its frequency and stability. We take an onsite-nonlinearity with a simple cubic form, but find that the nature of nonlinearity, e.g., softening or stiffening, greatly affects the linear stability of the nonlinear edge state.
To identify the type of the instabilities, we investigate the amplitude and phase of Floquet multipliers (FMs) in detail and employ a Krein signature analysis. Consequently,  we observe that high-amplitude edge states are generally unstable due to various types of instabilities caused by the collisions among isolated (localized) and band (extended) FMs. Therefore, an initially-excited nonlinear-edge state loses its energy and moves to the stable linear edge state over a long time.

However, for stiffening nonlinearity, we find a frequency-regime where high-amplitude edge states do \textit{not} show any instability. This happens when spectral bands of FMs are well separated enough to avoid any intersection and do not interact with any localized bifurcating state. This therefore opens the possibility of localizing a large amount of energy on the boundaries of the system for a long time. 

By adding disorder to our nonlinear system, we find that the frequency and stability of the nonlinear edge state show lesser deviation, in general, under a chiral disorder compared to a non-chiral disorder. Interestingly, we also find that in the presence of weak disorder (5\%) of both types, in the stiffening case, there is still a large frequency regime where high-amplitude edge states can be linearly stable. This stability region is larger for the case with chiral disorder compared to non-chiral disorder, which we explain by estimating the shifts in cutoff frequencies due to disorder.

A main finding of the work is that it is possible, under suitable nonlinear conditions, for topological linear states to remain robust. While onsite nonlinearity may introduce instabilities, the parametric range of instability and its type depends on the details of nonlinearity and system parameters. 
This suggests that it will be interesting to investigate the role of inter-site nonlinearity on the stability. In addition, functional forms of nonlinearity other than cubic could be investigated. Given that mechanical systems offer extreme tunability of nonlinear responses, it will be interesting to come up with the required nature of nonlinearity, for example, by topological optimization techniques, that maximizes the range of frequencies where high-amplitude topological edge states are stable. Finally, this study can inspire a systematic exploration of stability of topological corner, edge, and surface states in higher-dimensional nonlinear systems. There, the topological nature of the states manifests itself in a variety of ways, such as their ability to potentially bypass impurities and transmit over domain boundaries without losing energy to the bulk of the domain. In that setting, another aspect related to the mobility of nonlinear states comes into play, a topic that has been extensively studied~\cite{Marin1998b}.

\section*{Acknowledgments}
R. C. thanks Dr. Hiromi Yasuda (University of Pennsylvania, USA) for the help in Python programming.  R. C. and G. T. acknowledge the support from the project CS.MICRO funded under the program Etoiles Montantes of the Region Pays de la Loire, France. H. X. acknowledges the support from NSFC (Grant No. 11801191). J. Y. is grateful for the support from NSF (CAREER-1553202 and EFRI-1741685). P. G. K. acknowledges the support from NSF (DMS-1809074).

\appendix
\section{Floquet theory for linear stability} 
\label{Appendix_A}
For the sake of completeness, in this appendix we review Floquet theory for the linear stability of a nonlinear time-periodic state. Since we consider a 1D lattice with $n$ masses connected with springs, its Newtonian dynamics can be analyzed by taking a state vector $\mathbf{q} =(\mathbf{x} , \dot{\mathbf{x} })=(x_1, x_2,...x_n, \dot{x}_1, \dot{x}_2,...\dot{x}_n)$, where $x_j$ and $\dot{x}_j$ denote the displacement and velocity of the $j$th mass. We thus write a set of $2n$ first-order ODEs as a function of $\mathbf{q}$

\begin{eqnarray}
\dot{\mathbf{q} }  = F(\mathbf{q} ).
\end{eqnarray}

\noindent We then use Newton's method to find a periodic solution $\mathbf{q}_0 =(\mathbf{x}_0 , \dot{\mathbf{x}}_0 )$ for a given time period $T$. These are the nonlinear solutions we are looking for in this study.

Linear stability of such nonlinear solutions is determined by tracking the evolution of a small perturbation $\mathbf{dq}$ on the periodic orbit $\mathbf{q}_0 $. Therefore, by substituting $\mathbf{q} = \mathbf{q}_0 + \mathbf{dq} $, we obtained the system of variational equations:

\begin{eqnarray}
\dot{\mathbf{dq} }  = \left(\frac{\partial F}{\partial \mathbf{q}} \right)_{\mathbf{q}=\mathbf{q}_0}   \mathbf{dq}.
\label{eq:Floq}
\end{eqnarray}

\noindent These are linear differential equations with time-periodic coefficients. Therefore, the general solution, i.e., the perturbation at time $t$, is given by Floquet theory as
\begin{eqnarray}
\mathbf{dq}(t) =\mathrm{\Delta}(t) \mathbf{dq} (0),
\end{eqnarray}

\noindent where $\mathrm{\Delta}(t)$ and $\mathbf{dq} (0)$ are the \textit{fundamental matrix} of solutions and initial perturbation, respectively. 

Next, the perturbation after $t=T$ can be easily written as
\begin{eqnarray}
\mathbf{dq}(T) =\mathrm{\Delta}(T) \mathbf{dq} (0),
\end{eqnarray}
\noindent where $\mathrm{\Delta}(T)$ is called the \textit{monodromy matrix}. We can further generalize this to calculate the perturbation after $m$ time periods as
\begin{eqnarray}
\mathbf{dq}(m T) =\left [ \mathrm{\Delta}(T)  \right]^m \mathbf{dq} (0).
\label{eq:map}
\end{eqnarray}

Clearly, for linear stability of the periodic solution $\mathbf{q}_0$, we are interested in knowing how the small perturbation $\mathbf{dq} (0)$ grows over time. Therefore Eq. \eqref{eq:map} is the key mapping, and the eigenvalues of the monodromy matrix $\mathrm{\Delta}(T) $, known as Floquet multipliers (FMs), determine the linear stability of the periodic orbit. Since the monodromy matrix is real, the eigenvalues come in complex conjugate pairs.

Let $\lambda$ be a FM, then the perturbation evolves as 
\begin{eqnarray}
\mathbf{dq}(t) = g(t) \lambda^{t/T},
\label{eq:lambda}
\end{eqnarray}

\noindent where $g(t)$ is $T$-periodic and the initial value $g(0)$ is the eigenvector $v$ corresponding to the eigenvalue $\lambda$. 
Clearly, the perturbation grows exponentially over time when $\abs{\lambda}>1$, while for $\abs{\lambda} \le 1$ the dynamics is linearly stable. Since our system is a Hamiltonian system, its FMs multipliers comes in complex conjugate and reciprocal pairs. This means that for a complex FM $\lambda$, we also have $1/\lambda$, $\lambda^*$, and $1/\lambda^*$ as other FMs, where $*$ denotes complex conjugation. We, therefore, conclude that the FMs must lie on the unit circle ($\abs{\lambda} = 1$) for linear stability in our system and any deviation from unity represents the existence of instability.
When $\abs{\lambda}$ deviates from unity, the relative growth of the initial state after $m$ time periods can be estimated as
\begin{eqnarray}
\frac{\abs{\mathbf{dq}(m T)} - \abs{\mathbf{dq}(0)} } {\abs{\mathbf{dq}(0)}}  &=& \abs{\lambda}^{m} -1, \nonumber \\
&\approx& m (\abs{\lambda}-1) \ \ \text{if} \ \ \abs{\lambda} \approx 1.
\label{eq:growth}
\end{eqnarray}

\section{Spectral bands in FM phase plots}
\label{Appendix_B}
 \begin{figure}[!]
\centering
\includegraphics[width=\columnwidth]{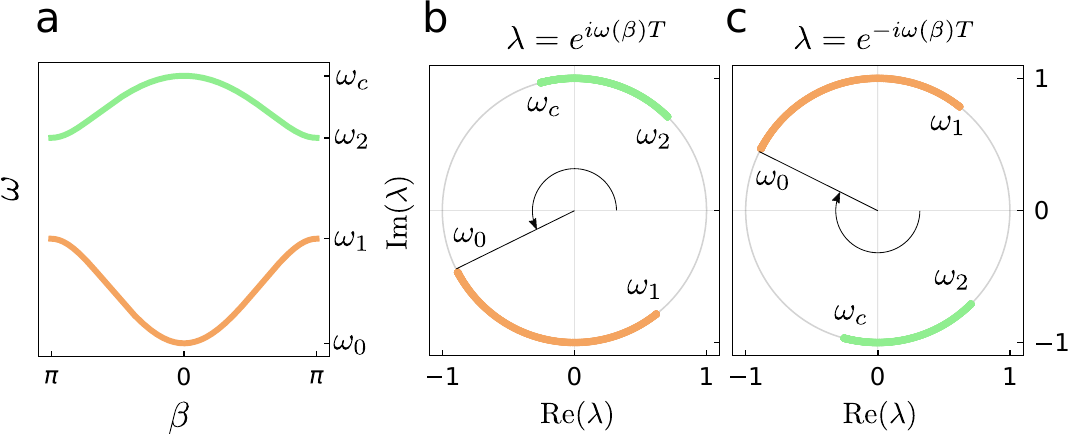}
\caption{Mapping between the dispersion and the corresponding FMs. (a) Dispersion for our linearized system with $\gamma=0.4$ and $\gamma_0=1$. The cutoff frequencies are marked for the acoustic (below) and optical (above) bands. (b) Corresponding FMs $\lambda = e^{i \omega(\beta) T}$ on the complex plane for $T=2 \pi/\omega_{ed}$. Edges of the arcs are marked with the associated cutoff frequency of dispersion band. (c)  Corresponding FM $\lambda = e^{-i \omega(\beta) T}$ for the same $T$.}
\label{fig7}
\end{figure} 

We observed spectral bands in Figs.~\ref{fig4}c,f of the main text. In this appendix, we elaborate on their connection to the dispersion bands (acoustic and optical) of the linearized system.
In the process to obtain FMs, we perform linearization over the shape of the nonlinear edge state, in the far-field limit (away from the localized edge) where the solution decays to zero. 
This vanishing background supports linear dynamics since the system is linearizable at zero amplitude. 
Let the infinite length of such background follow the dispersion $\omega(\beta)$, where $\beta$ is the normalized wave vector. Therefore, a perturbation $\mathbf{dq}(t)$ along the eigenvector $g(0)$ evolves in time as
\begin{eqnarray}
\mathbf{dq}(t) = g(0) e^{\pm i \omega(\beta) t},
\end{eqnarray}

\noindent which can be recast as
\begin{eqnarray}
\mathbf{dq}(t) = g(0) \left[ e^{\pm i \omega(\beta) T} \right]^{t/T}.
\end{eqnarray}

\noindent By comparing this equation to the more general Eq.~\eqref{eq:lambda}, we deduce  
\begin{eqnarray}
\lambda = e^{\pm i \omega(\beta) T}.
\label{eq:lambdaT}
\end{eqnarray}

\noindent Therefore, we conclude that the FMs obtained from the vanishing background (far away from the edge) lie on the unit circle in the complex plane. Their phase is dependent on the dispersion $\omega(\beta)$ of the linearized system and the time period $T$ of the background oscillation of the edge state.

In Fig.~\ref{fig7}a, we show the dispersion $\omega(\beta)$ for our system with $\gamma=0.4$ and $\gamma_0=1$. 
The lower band (acoustic) starts at frequency $\omega_0=\sqrt{\gamma_0}$ and ends at $\omega_1 = \sqrt{2(1-\gamma) + \gamma_0}$. The upper band (optical) starts at $\omega_2 = \sqrt{2(1 + \gamma) + \gamma_0}$ and ends at $\omega_c = \sqrt{4 + \gamma_0}$. Using Eq.~\eqref{eq:lambdaT} we map these bands to the FMs on the complex plane for $T=2\pi/\omega_{ed}$ and show them in Figs.~\ref{fig7}b,c.  The dispersion bands are mapped as arcs on the complex plane. We get two complex conjugate pairs of these, shown separately in Figs.~\ref{fig7}b,c when we take different signs in Eq.~\eqref{eq:lambdaT}. The actual mapping is the superposition of these two. This, therefore, explains the origin of spectral bands when FM phase is plotted in Figs.~\ref{fig4}c,f.

Now we can easily write a general expression for the span of these arcs on the unit circle for the nonlinear edge state at frequency $\omega$.
FM phase thus vary in the interval:
\begin{eqnarray}
\left. \mathrm{arg} (\lambda)\right|_{\mathrm{acoustic}} &\in & \left[ \frac{2 \pi \omega_0}{\omega} -2 \pi p,   \frac{2 \pi \omega_1}{\omega} -2 \pi p \right]  \label{eq:ph1},\\
\left. \mathrm{arg} (\lambda)\right|_{\mathrm{optical}}  &\in & \left[ \frac{2 \pi \omega_2}{\omega} -2 \pi p,  \frac{2 \pi \omega_c}{\omega} -2 \pi p \right]\label{eq:ph2},
\end{eqnarray}

\noindent where $p$ is an integer to bring the phases in the range of $[-\pi , \pi]$. In the upper half of Figs.~\ref{fig4}c,f, the band with negative Krein signature (orange) thus corresponds to the acoustic branch of the dispersion, whereas the band with positive Krein signature (green) corresponds to the optical branch. 

We can use the aforementioned expression to analytically estimate the region of linear stability shown in Fig.~\ref{fig4}f. It is important to note that we establish only a necessary condition to have no ``bulk-bulk'' instability. This is not a sufficient condition since our analytics do not predict the onset of bifurcations, and thus, can not predict the ``bulk-edge'' instabilities. For the set of parameters ($\gamma=0.4$ and $\gamma_0=1$), the ``bulk-bulk'' instabilities do not exist after the point where the cutoff phases of acoustic and optical bands intersect (circle in Fig.~\ref{fig4}f). The FM phases are $2 \pi (1 - \omega_1/\omega)$ and $2 \pi (\omega_c/\omega -1)$, respectively, for the acoustic and optical band cutoffs. For the region without   the``bulk-bulk'' instabilities, we must have
\begin{eqnarray}
2 \pi (1 - \omega_1/\omega) &>& 2 \pi (\omega_c/\omega -1), \nonumber\\
\Rightarrow \omega &>&\frac{\omega_1 + \omega_c}{2} = 1.86,
\end{eqnarray}
\noindent which is consistent with our observation in Fig.~\ref{fig4}f. 

\section{Effect of disorder on the stability of the nonlinear edge state}
\label{Appendix_C}
 \begin{figure}[!]
\centering
\includegraphics[width=\columnwidth]{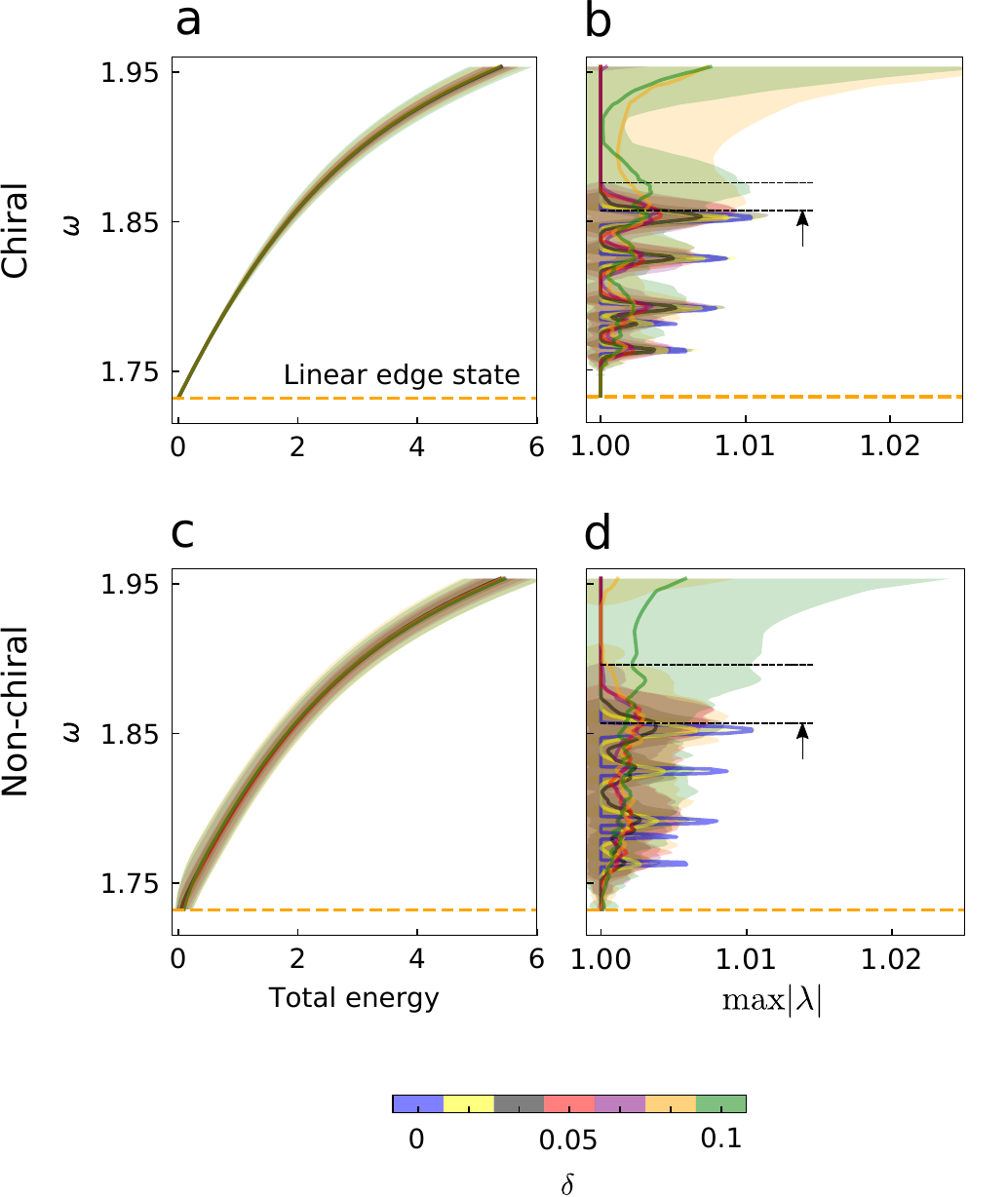}
\caption{Effect of disorder on the nonlinear continuation and instabilities of the edge state for stiffening nonlinearity. The conditions are similar to the ones in Figs. \ref{fig6}e-h but for more steps in disorder strength and 100 realizations of each. A short chain of 10 particles in considered.}
\label{fig8}
\end{figure} 
In this appendix, we further support our conjecture that the presence of weak disorder \textit{gradually} reduces the frequency span where nonlinear edge state are linearly stable. Recall that the calculations in Fig.~\ref{fig6} of the main text are carried out for only two, 5\% and 10\% disorder strengths, each having 50 realizations. Here, we take 6 different strengths of disorder till 10\% with 100 realizations of each. We consider a shorter chain (10 particles) for fast computation and analyze the case of stiffening nonlinearity as shown in Figs. \ref{fig6}e-h. In Fig.~\ref{fig8}, we show the similar trends. First, non-chiral disorder leads to more variation in energy and instability, mainly in the low-frequency regime. Second, the frequency range of stability gradually reduces with the increase of disorder strength and remains intact till about $6.6\%$. The reduction is large in the case of non-chiral disorder as shown with dashed lines. Third, for even higher disorder, i.e., 10\%, we observe the onset of instabilities in both cases, resulting in the large variation in FMs. This is dominated by the change in bifurcation and collision of isolated FMs as discussed in the main text.
This suggests that for weak disorder, the stability range for the nonlinear edge state is gradually reduced up to a disorder threshold and a quantitative estimate (conservative) of this reduction can be made from analyzing the linear spectrum as done in Fig.~\ref{fig5}.

\def\bibsection{\section*{References}} 

\bibliographystyle{apsrev4-1}

\bibliography{references.bib}

\begin{thebibliography}{64}%
\makeatletter
\providecommand \@ifxundefined [1]{%
 \@ifx{#1\undefined}
}%
\providecommand \@ifnum [1]{%
 \ifnum #1\expandafter \@firstoftwo
 \else \expandafter \@secondoftwo
 \fi
}%
\providecommand \@ifx [1]{%
 \ifx #1\expandafter \@firstoftwo
 \else \expandafter \@secondoftwo
 \fi
}%
\providecommand \natexlab [1]{#1}%
\providecommand \enquote  [1]{``#1''}%
\providecommand \bibnamefont  [1]{#1}%
\providecommand \bibfnamefont [1]{#1}%
\providecommand \citenamefont [1]{#1}%
\providecommand \href@noop [0]{\@secondoftwo}%
\providecommand \href [0]{\begingroup \@sanitize@url \@href}%
\providecommand \@href[1]{\@@startlink{#1}\@@href}%
\providecommand \@@href[1]{\endgroup#1\@@endlink}%
\providecommand \@sanitize@url [0]{\catcode `\\12\catcode `\$12\catcode
  `\&12\catcode `\#12\catcode `\^12\catcode `\_12\catcode `\%12\relax}%
\providecommand \@@startlink[1]{}%
\providecommand \@@endlink[0]{}%
\providecommand \url  [0]{\begingroup\@sanitize@url \@url }%
\providecommand \@url [1]{\endgroup\@href {#1}{\urlprefix }}%
\providecommand \urlprefix  [0]{URL }%
\providecommand \Eprint [0]{\href }%
\providecommand \doibase [0]{http://dx.doi.org/}%
\providecommand \selectlanguage [0]{\@gobble}%
\providecommand \bibinfo  [0]{\@secondoftwo}%
\providecommand \bibfield  [0]{\@secondoftwo}%
\providecommand \translation [1]{[#1]}%
\providecommand \BibitemOpen [0]{}%
\providecommand \bibitemStop [0]{}%
\providecommand \bibitemNoStop [0]{.\EOS\space}%
\providecommand \EOS [0]{\spacefactor3000\relax}%
\providecommand \BibitemShut  [1]{\csname bibitem#1\endcsname}%
\let\auto@bib@innerbib\@empty
\bibitem [{\citenamefont {Hasan}\ and\ \citenamefont {Kane}(2010)}]{Hasan2010}%
  \BibitemOpen
  \bibfield  {author} {\bibinfo {author} {\bibfnamefont {M.~Z.}\ \bibnamefont
  {Hasan}}\ and\ \bibinfo {author} {\bibfnamefont {C.~L.}\ \bibnamefont
  {Kane}},\ }\bibfield  {title} {\emph {\bibinfo {title} {{Colloquium :
  Topological insulators}},\ }}\href {\doibase 10.1103/RevModPhys.82.3045}
  {\bibfield  {journal} {\bibinfo  {journal} {Rev. Mod. Phys.}\ }\textbf
  {\bibinfo {volume} {82}},\ \bibinfo {pages} {3045} (\bibinfo {year}
  {2010})}\BibitemShut {NoStop}%
\bibitem [{\citenamefont {Cooper}\ \emph {et~al.}(2019)\citenamefont {Cooper},
  \citenamefont {Dalibard},\ and\ \citenamefont {Spielman}}]{Cooper2019}%
  \BibitemOpen
  \bibfield  {author} {\bibinfo {author} {\bibfnamefont {N.~R.}\ \bibnamefont
  {Cooper}}, \bibinfo {author} {\bibfnamefont {J.}~\bibnamefont {Dalibard}}, \
  and\ \bibinfo {author} {\bibfnamefont {I.~B.}\ \bibnamefont {Spielman}},\
  }\bibfield  {title} {\emph {\bibinfo {title} {Topological bands for ultracold
  atoms},\ }}\href {\doibase 10.1103/RevModPhys.91.015005} {\bibfield
  {journal} {\bibinfo  {journal} {Rev. Mod. Phys.}\ }\textbf {\bibinfo {volume}
  {91}},\ \bibinfo {pages} {015005} (\bibinfo {year} {2019})}\BibitemShut
  {NoStop}%
\bibitem [{\citenamefont {Ozawa}\ \emph {et~al.}(2019)\citenamefont {Ozawa},
  \citenamefont {Price}, \citenamefont {Amo}, \citenamefont {Goldman},
  \citenamefont {Hafezi}, \citenamefont {Lu}, \citenamefont {Rechtsman},
  \citenamefont {Schuster}, \citenamefont {Simon}, \citenamefont {Zilberberg},\
  and\ \citenamefont {Carusotto}}]{Ozawa2019}%
  \BibitemOpen
  \bibfield  {author} {\bibinfo {author} {\bibfnamefont {T.}~\bibnamefont
  {Ozawa}}, \bibinfo {author} {\bibfnamefont {H.~M.}\ \bibnamefont {Price}},
  \bibinfo {author} {\bibfnamefont {A.}~\bibnamefont {Amo}}, \bibinfo {author}
  {\bibfnamefont {N.}~\bibnamefont {Goldman}}, \bibinfo {author} {\bibfnamefont
  {M.}~\bibnamefont {Hafezi}}, \bibinfo {author} {\bibfnamefont
  {L.}~\bibnamefont {Lu}}, \bibinfo {author} {\bibfnamefont {M.~C.}\
  \bibnamefont {Rechtsman}}, \bibinfo {author} {\bibfnamefont {D.}~\bibnamefont
  {Schuster}}, \bibinfo {author} {\bibfnamefont {J.}~\bibnamefont {Simon}},
  \bibinfo {author} {\bibfnamefont {O.}~\bibnamefont {Zilberberg}}, \ and\
  \bibinfo {author} {\bibfnamefont {I.}~\bibnamefont {Carusotto}},\ }\bibfield
  {title} {\emph {\bibinfo {title} {{Topological photonics}},\ }}\href
  {\doibase 10.1103/RevModPhys.91.015006} {\bibfield  {journal} {\bibinfo
  {journal} {Rev. Mod. Phys.}\ }\textbf {\bibinfo {volume} {91}},\ \bibinfo
  {pages} {015006} (\bibinfo {year} {2019})}\BibitemShut {NoStop}%
\bibitem [{\citenamefont {S{\"{u}}sstrunk}\ and\ \citenamefont
  {Huber}(2016)}]{Susstrunk2016}%
  \BibitemOpen
  \bibfield  {author} {\bibinfo {author} {\bibfnamefont {R.}~\bibnamefont
  {S{\"{u}}sstrunk}}\ and\ \bibinfo {author} {\bibfnamefont {S.~D.}\
  \bibnamefont {Huber}},\ }\bibfield  {title} {\emph {\bibinfo {title}
  {{Classification of topological phonons in linear mechanical
  metamaterials}},\ }}\href {\doibase 10.1073/pnas.1605462113} {\bibfield
  {journal} {\bibinfo  {journal} {Proc. Natl. Acad. Sci. USA}\ }\textbf
  {\bibinfo {volume} {113}},\ \bibinfo {pages} {E4767} (\bibinfo {year}
  {2016})}\BibitemShut {NoStop}%
\bibitem [{\citenamefont {Ma}\ \emph {et~al.}(2019)\citenamefont {Ma},
  \citenamefont {Xiao},\ and\ \citenamefont {Chan}}]{Ma2019}%
  \BibitemOpen
  \bibfield  {author} {\bibinfo {author} {\bibfnamefont {G.}~\bibnamefont
  {Ma}}, \bibinfo {author} {\bibfnamefont {M.}~\bibnamefont {Xiao}}, \ and\
  \bibinfo {author} {\bibfnamefont {C.~T.}\ \bibnamefont {Chan}},\ }\bibfield
  {title} {\emph {\bibinfo {title} {{Topological phases in acoustic and
  mechanical systems}},\ }}\href
  {http://www.nature.com/articles/s42254-019-0030-x} {\bibfield  {journal}
  {\bibinfo  {journal} {Nat. Rev. Phys.}\ }\textbf {\bibinfo {volume} {1}},\
  \bibinfo {pages} {281} (\bibinfo {year} {2019})}\BibitemShut {NoStop}%
\bibitem [{\citenamefont {Bernevig}\ and\ \citenamefont
  {Hughes}(2013)}]{Bernevig2013}%
  \BibitemOpen
  \bibfield  {author} {\bibinfo {author} {\bibfnamefont {B.~A.}\ \bibnamefont
  {Bernevig}}\ and\ \bibinfo {author} {\bibfnamefont {T.~L.}\ \bibnamefont
  {Hughes}},\ }\href {http://www.jstor.org/stable/j.ctt19cc2gc} {\emph
  {\bibinfo {title} {Topological Insulators and Topological Superconductors}}}\
  (\bibinfo  {publisher} {Princeton University Press},\ \bibinfo {year}
  {2013})\BibitemShut {NoStop}%
\bibitem [{\citenamefont {Kane}\ and\ \citenamefont {Mele}(2005)}]{Kane2005}%
  \BibitemOpen
  \bibfield  {author} {\bibinfo {author} {\bibfnamefont {C.~L.}\ \bibnamefont
  {Kane}}\ and\ \bibinfo {author} {\bibfnamefont {E.~J.}\ \bibnamefont
  {Mele}},\ }\bibfield  {title} {\emph {\bibinfo {title} {Quantum spin hall
  effect in graphene},\ }}\href {\doibase 10.1103/PhysRevLett.95.226801}
  {\bibfield  {journal} {\bibinfo  {journal} {Phys. Rev. Lett.}\ }\textbf
  {\bibinfo {volume} {95}},\ \bibinfo {pages} {226801} (\bibinfo {year}
  {2005})}\BibitemShut {NoStop}%
\bibitem [{\citenamefont {Wan}\ \emph {et~al.}(2011)\citenamefont {Wan},
  \citenamefont {Turner}, \citenamefont {Vishwanath},\ and\ \citenamefont
  {Savrasov}}]{Wan2011}%
  \BibitemOpen
  \bibfield  {author} {\bibinfo {author} {\bibfnamefont {X.}~\bibnamefont
  {Wan}}, \bibinfo {author} {\bibfnamefont {A.~M.}\ \bibnamefont {Turner}},
  \bibinfo {author} {\bibfnamefont {A.}~\bibnamefont {Vishwanath}}, \ and\
  \bibinfo {author} {\bibfnamefont {S.~Y.}\ \bibnamefont {Savrasov}},\
  }\bibfield  {title} {\emph {\bibinfo {title} {Topological semimetal and
  fermi-arc surface states in the electronic structure of pyrochlore
  iridates},\ }}\href {\doibase 10.1103/PhysRevB.83.205101} {\bibfield
  {journal} {\bibinfo  {journal} {Phys. Rev. B}\ }\textbf {\bibinfo {volume}
  {83}},\ \bibinfo {pages} {205101} (\bibinfo {year} {2011})}\BibitemShut
  {NoStop}%
\bibitem [{\citenamefont {Benalcazar}\ \emph {et~al.}(2017)\citenamefont
  {Benalcazar}, \citenamefont {Bernevig},\ and\ \citenamefont
  {Hughes}}]{Benalcazar2017}%
  \BibitemOpen
  \bibfield  {author} {\bibinfo {author} {\bibfnamefont {W.~A.}\ \bibnamefont
  {Benalcazar}}, \bibinfo {author} {\bibfnamefont {B.~A.}\ \bibnamefont
  {Bernevig}}, \ and\ \bibinfo {author} {\bibfnamefont {T.~L.}\ \bibnamefont
  {Hughes}},\ }\bibfield  {title} {\emph {\bibinfo {title} {Quantized electric
  multipole insulators},\ }}\href {\doibase 10.1126/science.aah6442} {\bibfield
   {journal} {\bibinfo  {journal} {Science}\ }\textbf {\bibinfo {volume}
  {357}},\ \bibinfo {pages} {61} (\bibinfo {year} {2017})}\BibitemShut
  {NoStop}%
\bibitem [{\citenamefont {Smirnova}\ \emph {et~al.}(2020)\citenamefont
  {Smirnova}, \citenamefont {Leykam}, \citenamefont {Chong},\ and\
  \citenamefont {Kivshar}}]{Smirnova2020}%
  \BibitemOpen
  \bibfield  {author} {\bibinfo {author} {\bibfnamefont {D.}~\bibnamefont
  {Smirnova}}, \bibinfo {author} {\bibfnamefont {D.}~\bibnamefont {Leykam}},
  \bibinfo {author} {\bibfnamefont {Y.}~\bibnamefont {Chong}}, \ and\ \bibinfo
  {author} {\bibfnamefont {Y.}~\bibnamefont {Kivshar}},\ }\bibfield  {title}
  {\emph {\bibinfo {title} {{Nonlinear topological photonics}},\ }}\href
  {https://aip.scitation.org/doi/10.1063/1.5142397} {\bibfield  {journal}
  {\bibinfo  {journal} {Appl. Phys. Rev.}\ }\textbf {\bibinfo {volume} {7}},\
  \bibinfo {pages} {021306} (\bibinfo {year} {2020})}\BibitemShut {NoStop}%
\bibitem [{\citenamefont {Dobrykh}\ \emph {et~al.}(2018)\citenamefont
  {Dobrykh}, \citenamefont {Yulin}, \citenamefont {Slobozhanyuk}, \citenamefont
  {Poddubny},\ and\ \citenamefont {Kivshar}}]{Dobrykh2018}%
  \BibitemOpen
  \bibfield  {author} {\bibinfo {author} {\bibfnamefont {D.~A.}\ \bibnamefont
  {Dobrykh}}, \bibinfo {author} {\bibfnamefont {A.~V.}\ \bibnamefont {Yulin}},
  \bibinfo {author} {\bibfnamefont {A.~P.}\ \bibnamefont {Slobozhanyuk}},
  \bibinfo {author} {\bibfnamefont {A.~N.}\ \bibnamefont {Poddubny}}, \ and\
  \bibinfo {author} {\bibfnamefont {Y.~S.}\ \bibnamefont {Kivshar}},\
  }\bibfield  {title} {\emph {\bibinfo {title} {{Nonlinear Control of
  Electromagnetic Topological Edge States}},\ }}\href {\doibase
  10.1103/PhysRevLett.121.163901} {\bibfield  {journal} {\bibinfo  {journal}
  {Phys. Rev. Lett.}\ }\textbf {\bibinfo {volume} {121}},\ \bibinfo {pages}
  {163901} (\bibinfo {year} {2018})}\BibitemShut {NoStop}%
\bibitem [{\citenamefont {Pal}\ \emph {et~al.}(2018)\citenamefont {Pal},
  \citenamefont {Vila}, \citenamefont {Leamy},\ and\ \citenamefont
  {Ruzzene}}]{Pal2018}%
  \BibitemOpen
  \bibfield  {author} {\bibinfo {author} {\bibfnamefont {R.~K.}\ \bibnamefont
  {Pal}}, \bibinfo {author} {\bibfnamefont {J.}~\bibnamefont {Vila}}, \bibinfo
  {author} {\bibfnamefont {M.}~\bibnamefont {Leamy}}, \ and\ \bibinfo {author}
  {\bibfnamefont {M.}~\bibnamefont {Ruzzene}},\ }\bibfield  {title} {\emph
  {\bibinfo {title} {{Amplitude-dependent topological edge states in nonlinear
  phononic lattices}},\ }}\href {\doibase 10.1103/PhysRevE.97.032209}
  {\bibfield  {journal} {\bibinfo  {journal} {Phys. Rev. E}\ }\textbf {\bibinfo
  {volume} {97}},\ \bibinfo {pages} {032209} (\bibinfo {year}
  {2018})}\BibitemShut {NoStop}%
\bibitem [{\citenamefont {Vila}\ \emph {et~al.}(2019)\citenamefont {Vila},
  \citenamefont {Paulino},\ and\ \citenamefont {Ruzzene}}]{Vila2019}%
  \BibitemOpen
  \bibfield  {author} {\bibinfo {author} {\bibfnamefont {J.}~\bibnamefont
  {Vila}}, \bibinfo {author} {\bibfnamefont {G.~H.}\ \bibnamefont {Paulino}}, \
  and\ \bibinfo {author} {\bibfnamefont {M.}~\bibnamefont {Ruzzene}},\
  }\bibfield  {title} {\emph {\bibinfo {title} {Role of nonlinearities in
  topological protection: Testing magnetically coupled fidget spinners},\
  }}\href {\doibase 10.1103/PhysRevB.99.125116} {\bibfield  {journal} {\bibinfo
   {journal} {Phys. Rev. B}\ }\textbf {\bibinfo {volume} {99}},\ \bibinfo
  {pages} {125116} (\bibinfo {year} {2019})}\BibitemShut {NoStop}%
\bibitem [{\citenamefont {Kruk}\ \emph {et~al.}(2019)\citenamefont {Kruk},
  \citenamefont {Poddubny}, \citenamefont {Smirnova}, \citenamefont {Wang},
  \citenamefont {Slobozhanyuk}, \citenamefont {Shorokhov}, \citenamefont
  {Kravchenko}, \citenamefont {Luther-Davies},\ and\ \citenamefont
  {Kivshar}}]{Kruk2019}%
  \BibitemOpen
  \bibfield  {author} {\bibinfo {author} {\bibfnamefont {S.}~\bibnamefont
  {Kruk}}, \bibinfo {author} {\bibfnamefont {A.}~\bibnamefont {Poddubny}},
  \bibinfo {author} {\bibfnamefont {D.}~\bibnamefont {Smirnova}}, \bibinfo
  {author} {\bibfnamefont {L.}~\bibnamefont {Wang}}, \bibinfo {author}
  {\bibfnamefont {A.}~\bibnamefont {Slobozhanyuk}}, \bibinfo {author}
  {\bibfnamefont {A.}~\bibnamefont {Shorokhov}}, \bibinfo {author}
  {\bibfnamefont {I.}~\bibnamefont {Kravchenko}}, \bibinfo {author}
  {\bibfnamefont {B.}~\bibnamefont {Luther-Davies}}, \ and\ \bibinfo {author}
  {\bibfnamefont {Y.}~\bibnamefont {Kivshar}},\ }\bibfield  {title} {\emph
  {\bibinfo {title} {{Nonlinear light generation in topological
  nanostructures}},\ }}\href {\doibase 10.1038/s41565-018-0324-7} {\bibfield
  {journal} {\bibinfo  {journal} {Nat. Nanotechnol.}\ }\textbf {\bibinfo
  {volume} {14}},\ \bibinfo {pages} {126} (\bibinfo {year} {2019})}\BibitemShut
  {NoStop}%
\bibitem [{\citenamefont {Wang}\ \emph {et~al.}(2019)\citenamefont {Wang},
  \citenamefont {Lang}, \citenamefont {Lee}, \citenamefont {Zhang},\ and\
  \citenamefont {Chong}}]{Wang2019}%
  \BibitemOpen
  \bibfield  {author} {\bibinfo {author} {\bibfnamefont {Y.}~\bibnamefont
  {Wang}}, \bibinfo {author} {\bibfnamefont {L.-J.}\ \bibnamefont {Lang}},
  \bibinfo {author} {\bibfnamefont {C.~H.}\ \bibnamefont {Lee}}, \bibinfo
  {author} {\bibfnamefont {B.}~\bibnamefont {Zhang}}, \ and\ \bibinfo {author}
  {\bibfnamefont {Y.~D.}\ \bibnamefont {Chong}},\ }\bibfield  {title} {\emph
  {\bibinfo {title} {{Topologically enhanced harmonic generation in a nonlinear
  transmission line metamaterial}},\ }}\href {\doibase
  10.1038/s41467-019-08966-9} {\bibfield  {journal} {\bibinfo  {journal} {Nat.
  Commun.}\ }\textbf {\bibinfo {volume} {10}},\ \bibinfo {pages} {1102}
  (\bibinfo {year} {2019})}\BibitemShut {NoStop}%
\bibitem [{\citenamefont {Darabi}\ and\ \citenamefont
  {Leamy}(2019)}]{Darabi2019}%
  \BibitemOpen
  \bibfield  {author} {\bibinfo {author} {\bibfnamefont {A.}~\bibnamefont
  {Darabi}}\ and\ \bibinfo {author} {\bibfnamefont {M.~J.}\ \bibnamefont
  {Leamy}},\ }\bibfield  {title} {\emph {\bibinfo {title} {{Tunable Nonlinear
  Topological Insulator for Acoustic Waves}},\ }}\href {\doibase
  10.1103/PhysRevApplied.12.044030} {\bibfield  {journal} {\bibinfo  {journal}
  {Phys. Rev. Applied}\ }\textbf {\bibinfo {volume} {12}},\ \bibinfo {pages}
  {044030} (\bibinfo {year} {2019})}\BibitemShut {NoStop}%
\bibitem [{\citenamefont {Zhou}\ \emph {et~al.}(2020)\citenamefont {Zhou},
  \citenamefont {Ma}, \citenamefont {Sun}, \citenamefont {Gonella},\ and\
  \citenamefont {Mao}}]{Zhou2020}%
  \BibitemOpen
  \bibfield  {author} {\bibinfo {author} {\bibfnamefont {D.}~\bibnamefont
  {Zhou}}, \bibinfo {author} {\bibfnamefont {J.}~\bibnamefont {Ma}}, \bibinfo
  {author} {\bibfnamefont {K.}~\bibnamefont {Sun}}, \bibinfo {author}
  {\bibfnamefont {S.}~\bibnamefont {Gonella}}, \ and\ \bibinfo {author}
  {\bibfnamefont {X.}~\bibnamefont {Mao}},\ }\bibfield  {title} {\emph
  {\bibinfo {title} {{Switchable phonon diodes using nonlinear topological
  Maxwell lattices}},\ }}\href {\doibase 10.1103/PhysRevB.101.104106}
  {\bibfield  {journal} {\bibinfo  {journal} {Phys. Rev. B}\ }\textbf {\bibinfo
  {volume} {101}},\ \bibinfo {pages} {104106} (\bibinfo {year}
  {2020})}\BibitemShut {NoStop}%
\bibitem [{\citenamefont {Ablowitz}\ \emph {et~al.}(2014)\citenamefont
  {Ablowitz}, \citenamefont {Curtis},\ and\ \citenamefont {Ma}}]{Ablowitz2014}%
  \BibitemOpen
  \bibfield  {author} {\bibinfo {author} {\bibfnamefont {M.~J.}\ \bibnamefont
  {Ablowitz}}, \bibinfo {author} {\bibfnamefont {C.~W.}\ \bibnamefont
  {Curtis}}, \ and\ \bibinfo {author} {\bibfnamefont {Y.-P.}\ \bibnamefont
  {Ma}},\ }\bibfield  {title} {\emph {\bibinfo {title} {{Linear and nonlinear
  traveling edge waves in optical honeycomb lattices}},\ }}\href {\doibase
  10.1103/PhysRevA.90.023813} {\bibfield  {journal} {\bibinfo  {journal} {Phys.
  Rev. A}\ }\textbf {\bibinfo {volume} {90}},\ \bibinfo {pages} {023813}
  (\bibinfo {year} {2014})}\BibitemShut {NoStop}%
\bibitem [{\citenamefont {Leykam}\ and\ \citenamefont
  {Chong}(2016)}]{Leykam2016}%
  \BibitemOpen
  \bibfield  {author} {\bibinfo {author} {\bibfnamefont {D.}~\bibnamefont
  {Leykam}}\ and\ \bibinfo {author} {\bibfnamefont {Y.~D.}\ \bibnamefont
  {Chong}},\ }\bibfield  {title} {\emph {\bibinfo {title} {{Edge Solitons in
  Nonlinear-Photonic Topological Insulators}},\ }}\href {\doibase
  10.1103/PhysRevLett.117.143901} {\bibfield  {journal} {\bibinfo  {journal}
  {Phys. Rev. Lett.}\ }\textbf {\bibinfo {volume} {117}},\ \bibinfo {pages}
  {143901} (\bibinfo {year} {2016})}\BibitemShut {NoStop}%
\bibitem [{\citenamefont {Kartashov}\ and\ \citenamefont
  {Skryabin}(2016)}]{Kartashov2016}%
  \BibitemOpen
  \bibfield  {author} {\bibinfo {author} {\bibfnamefont {Y.~V.}\ \bibnamefont
  {Kartashov}}\ and\ \bibinfo {author} {\bibfnamefont {D.~V.}\ \bibnamefont
  {Skryabin}},\ }\bibfield  {title} {\emph {\bibinfo {title} {{Modulational
  instability and solitary waves in polariton topological insulators}},\
  }}\href {\doibase 10.1364/optica.3.001228} {\bibfield  {journal} {\bibinfo
  {journal} {Optica}\ }\textbf {\bibinfo {volume} {3}},\ \bibinfo {pages}
  {1228} (\bibinfo {year} {2016})}\BibitemShut {NoStop}%
\bibitem [{\citenamefont {Snee}\ and\ \citenamefont {Ma}(2019)}]{Snee2019}%
  \BibitemOpen
  \bibfield  {author} {\bibinfo {author} {\bibfnamefont {D.~D.}\ \bibnamefont
  {Snee}}\ and\ \bibinfo {author} {\bibfnamefont {Y.-P.}\ \bibnamefont {Ma}},\
  }\bibfield  {title} {\emph {\bibinfo {title} {Edge solitons in a nonlinear
  mechanical topological insulator},\ }}\href {\doibase
  https://doi.org/10.1016/j.eml.2019.100487} {\bibfield  {journal} {\bibinfo
  {journal} {Extreme Mech. Lett.}\ }\textbf {\bibinfo {volume} {30}},\ \bibinfo
  {pages} {100487} (\bibinfo {year} {2019})}\BibitemShut {NoStop}%
\bibitem [{\citenamefont {Tao}\ \emph {et~al.}(2020)\citenamefont {Tao},
  \citenamefont {Dai}, \citenamefont {Yang}, \citenamefont {Zeng},\ and\
  \citenamefont {Xu}}]{Tao2020}%
  \BibitemOpen
  \bibfield  {author} {\bibinfo {author} {\bibfnamefont {Y.-L.}\ \bibnamefont
  {Tao}}, \bibinfo {author} {\bibfnamefont {N.}~\bibnamefont {Dai}}, \bibinfo
  {author} {\bibfnamefont {Y.-B.}\ \bibnamefont {Yang}}, \bibinfo {author}
  {\bibfnamefont {Q.-B.}\ \bibnamefont {Zeng}}, \ and\ \bibinfo {author}
  {\bibfnamefont {Y.}~\bibnamefont {Xu}},\ }\bibfield  {title} {\emph {\bibinfo
  {title} {{Hinge solitons in three-dimensional second-order topological
  insulators}},\ }}\href {http://arxiv.org/abs/2005.04433} {\bibfield
  {journal} {\bibinfo  {journal} {arXiv:2005.04433}\ } (\bibinfo {year}
  {2020})}\BibitemShut {NoStop}%
\bibitem [{\citenamefont {Mukherjee}\ and\ \citenamefont
  {Rechtsman}(2020{\natexlab{a}})}]{Mukherjee2020b}%
  \BibitemOpen
  \bibfield  {author} {\bibinfo {author} {\bibfnamefont {S.}~\bibnamefont
  {Mukherjee}}\ and\ \bibinfo {author} {\bibfnamefont {M.~C.}\ \bibnamefont
  {Rechtsman}},\ }\bibfield  {title} {\emph {\bibinfo {title} {{Observation of
  unidirectional soliton-like edge states in nonlinear Floquet topological
  insulators}},\ }}\href {https://arxiv.org/abs/2010.11359} {\bibfield
  {journal} {\bibinfo  {journal} {arXiv:2020.11359}\ } (\bibinfo {year}
  {2020}{\natexlab{a}})}\BibitemShut {NoStop}%
\bibitem [{\citenamefont {Lumer}\ \emph {et~al.}(2013)\citenamefont {Lumer},
  \citenamefont {Plotnik}, \citenamefont {Rechtsman},\ and\ \citenamefont
  {Segev}}]{Lumer2013}%
  \BibitemOpen
  \bibfield  {author} {\bibinfo {author} {\bibfnamefont {Y.}~\bibnamefont
  {Lumer}}, \bibinfo {author} {\bibfnamefont {Y.}~\bibnamefont {Plotnik}},
  \bibinfo {author} {\bibfnamefont {M.~C.}\ \bibnamefont {Rechtsman}}, \ and\
  \bibinfo {author} {\bibfnamefont {M.}~\bibnamefont {Segev}},\ }\bibfield
  {title} {\emph {\bibinfo {title} {{Self-Localized States in Photonic
  Topological Insulators}},\ }}\href {\doibase 10.1103/PhysRevLett.111.243905}
  {\bibfield  {journal} {\bibinfo  {journal} {Phys. Rev. Lett.}\ }\textbf
  {\bibinfo {volume} {111}},\ \bibinfo {pages} {243905} (\bibinfo {year}
  {2013})}\BibitemShut {NoStop}%
\bibitem [{\citenamefont {Solnyshkov}\ \emph {et~al.}(2017)\citenamefont
  {Solnyshkov}, \citenamefont {Bleu}, \citenamefont {Teklu},\ and\
  \citenamefont {Malpuech}}]{Solnyshkov2017}%
  \BibitemOpen
  \bibfield  {author} {\bibinfo {author} {\bibfnamefont {D.~D.}\ \bibnamefont
  {Solnyshkov}}, \bibinfo {author} {\bibfnamefont {O.}~\bibnamefont {Bleu}},
  \bibinfo {author} {\bibfnamefont {B.}~\bibnamefont {Teklu}}, \ and\ \bibinfo
  {author} {\bibfnamefont {G.}~\bibnamefont {Malpuech}},\ }\bibfield  {title}
  {\emph {\bibinfo {title} {{Chirality of Topological Gap Solitons in Bosonic
  Dimer Chains}},\ }}\href {\doibase 10.1103/PhysRevLett.118.023901} {\bibfield
   {journal} {\bibinfo  {journal} {Phys. Rev. Lett.}\ }\textbf {\bibinfo
  {volume} {118}},\ \bibinfo {pages} {023901} (\bibinfo {year}
  {2017})}\BibitemShut {NoStop}%
\bibitem [{\citenamefont {Smirnova}\ \emph {et~al.}(2019)\citenamefont
  {Smirnova}, \citenamefont {Smirnov}, \citenamefont {Leykam},\ and\
  \citenamefont {Kivshar}}]{Smirnova2019}%
  \BibitemOpen
  \bibfield  {author} {\bibinfo {author} {\bibfnamefont {D.~A.}\ \bibnamefont
  {Smirnova}}, \bibinfo {author} {\bibfnamefont {L.~A.}\ \bibnamefont
  {Smirnov}}, \bibinfo {author} {\bibfnamefont {D.}~\bibnamefont {Leykam}}, \
  and\ \bibinfo {author} {\bibfnamefont {Y.~S.}\ \bibnamefont {Kivshar}},\
  }\bibfield  {title} {\emph {\bibinfo {title} {{Topological Edge States and
  Gap Solitons in the Nonlinear Dirac Model}},\ }}\href {\doibase
  10.1002/lpor.201900223} {\bibfield  {journal} {\bibinfo  {journal} {Laser
  Photonics Rev.}\ }\textbf {\bibinfo {volume} {13}},\ \bibinfo {pages}
  {1900223} (\bibinfo {year} {2019})}\BibitemShut {NoStop}%
\bibitem [{\citenamefont {Marzuola}\ \emph {et~al.}(2019)\citenamefont
  {Marzuola}, \citenamefont {Rechtsman}, \citenamefont {Osting},\ and\
  \citenamefont {Bandres}}]{Marzuola2019}%
  \BibitemOpen
  \bibfield  {author} {\bibinfo {author} {\bibfnamefont {J.~L.}\ \bibnamefont
  {Marzuola}}, \bibinfo {author} {\bibfnamefont {M.}~\bibnamefont {Rechtsman}},
  \bibinfo {author} {\bibfnamefont {B.}~\bibnamefont {Osting}}, \ and\ \bibinfo
  {author} {\bibfnamefont {M.}~\bibnamefont {Bandres}},\ }\bibfield  {title}
  {\emph {\bibinfo {title} {{Bulk soliton dynamics in bosonic topological
  insulators}},\ }}\href {http://arxiv.org/abs/1904.10312} {\bibfield
  {journal} {\bibinfo  {journal} {arXiv:1904.10312}\ } (\bibinfo {year}
  {2019})}\BibitemShut {NoStop}%
\bibitem [{\citenamefont {Mukherjee}\ and\ \citenamefont
  {Rechtsman}(2020{\natexlab{b}})}]{Mukherjee2020}%
  \BibitemOpen
  \bibfield  {author} {\bibinfo {author} {\bibfnamefont {S.}~\bibnamefont
  {Mukherjee}}\ and\ \bibinfo {author} {\bibfnamefont {M.~C.}\ \bibnamefont
  {Rechtsman}},\ }\bibfield  {title} {\emph {\bibinfo {title} {{Observation of
  Floquet solitons in a topological bandgap}},\ }}\href {\doibase
  10.1126/science.aba8725} {\bibfield  {journal} {\bibinfo  {journal}
  {Science}\ }\textbf {\bibinfo {volume} {368}},\ \bibinfo {pages} {856}
  (\bibinfo {year} {2020}{\natexlab{b}})}\BibitemShut {NoStop}%
\bibitem [{\citenamefont {Bomantara}\ \emph {et~al.}(2017)\citenamefont
  {Bomantara}, \citenamefont {Zhao}, \citenamefont {Zhou},\ and\ \citenamefont
  {Gong}}]{Bomantara2017}%
  \BibitemOpen
  \bibfield  {author} {\bibinfo {author} {\bibfnamefont {R.~W.}\ \bibnamefont
  {Bomantara}}, \bibinfo {author} {\bibfnamefont {W.}~\bibnamefont {Zhao}},
  \bibinfo {author} {\bibfnamefont {L.}~\bibnamefont {Zhou}}, \ and\ \bibinfo
  {author} {\bibfnamefont {J.}~\bibnamefont {Gong}},\ }\bibfield  {title}
  {\emph {\bibinfo {title} {Nonlinear dirac cones},\ }}\href {\doibase
  10.1103/PhysRevB.96.121406} {\bibfield  {journal} {\bibinfo  {journal} {Phys.
  Rev. B}\ }\textbf {\bibinfo {volume} {96}},\ \bibinfo {pages} {121406(R)}
  (\bibinfo {year} {2017})}\BibitemShut {NoStop}%
\bibitem [{\citenamefont {Hadad}\ \emph {et~al.}(2016)\citenamefont {Hadad},
  \citenamefont {Khanikaev},\ and\ \citenamefont {Al{\`{u}}}}]{Hadad2016}%
  \BibitemOpen
  \bibfield  {author} {\bibinfo {author} {\bibfnamefont {Y.}~\bibnamefont
  {Hadad}}, \bibinfo {author} {\bibfnamefont {A.~B.}\ \bibnamefont
  {Khanikaev}}, \ and\ \bibinfo {author} {\bibfnamefont {A.}~\bibnamefont
  {Al{\`{u}}}},\ }\bibfield  {title} {\emph {\bibinfo {title} {{Self-induced
  topological transitions and edge states supported by nonlinear staggered
  potentials}},\ }}\href {\doibase 10.1103/PhysRevB.93.155112} {\bibfield
  {journal} {\bibinfo  {journal} {Phys. Rev. B}\ }\textbf {\bibinfo {volume}
  {93}},\ \bibinfo {pages} {155112} (\bibinfo {year} {2016})}\BibitemShut
  {NoStop}%
\bibitem [{\citenamefont {Hadad}\ \emph {et~al.}(2018)\citenamefont {Hadad},
  \citenamefont {Soric}, \citenamefont {Khanikaev},\ and\ \citenamefont
  {Al{\`{u}}}}]{Hadad2018a}%
  \BibitemOpen
  \bibfield  {author} {\bibinfo {author} {\bibfnamefont {Y.}~\bibnamefont
  {Hadad}}, \bibinfo {author} {\bibfnamefont {J.~C.}\ \bibnamefont {Soric}},
  \bibinfo {author} {\bibfnamefont {A.~B.}\ \bibnamefont {Khanikaev}}, \ and\
  \bibinfo {author} {\bibfnamefont {A.}~\bibnamefont {Al{\`{u}}}},\ }\bibfield
  {title} {\emph {\bibinfo {title} {{Self-induced topological protection in
  nonlinear circuit arrays}},\ }}\href {\doibase 10.1038/s41928-018-0042-z}
  {\bibfield  {journal} {\bibinfo  {journal} {Nat. Electron.}\ }\textbf
  {\bibinfo {volume} {1}},\ \bibinfo {pages} {178} (\bibinfo {year}
  {2018})}\BibitemShut {NoStop}%
\bibitem [{\citenamefont {Savelev}\ \emph {et~al.}(2018)\citenamefont
  {Savelev}, \citenamefont {Gorlach},\ and\ \citenamefont
  {Poddubny}}]{Savelev2018}%
  \BibitemOpen
  \bibfield  {author} {\bibinfo {author} {\bibfnamefont {R.~S.}\ \bibnamefont
  {Savelev}}, \bibinfo {author} {\bibfnamefont {M.~A.}\ \bibnamefont
  {Gorlach}}, \ and\ \bibinfo {author} {\bibfnamefont {A.~N.}\ \bibnamefont
  {Poddubny}},\ }\bibfield  {title} {\emph {\bibinfo {title} {{Topological
  interface states mediated by spontaneous symmetry breaking}},\ }}\href
  {\doibase 10.1103/PhysRevB.98.045415} {\bibfield  {journal} {\bibinfo
  {journal} {Phys. Rev. B}\ }\textbf {\bibinfo {volume} {98}},\ \bibinfo
  {pages} {045415} (\bibinfo {year} {2018})}\BibitemShut {NoStop}%
\bibitem [{\citenamefont {Chaunsali}\ and\ \citenamefont
  {Theocharis}(2019)}]{Chaunsali2019}%
  \BibitemOpen
  \bibfield  {author} {\bibinfo {author} {\bibfnamefont {R.}~\bibnamefont
  {Chaunsali}}\ and\ \bibinfo {author} {\bibfnamefont {G.}~\bibnamefont
  {Theocharis}},\ }\bibfield  {title} {\emph {\bibinfo {title} {{Self-induced
  topological transition in phononic crystals by nonlinearity management}},\
  }}\href {\doibase 10.1103/PhysRevB.100.014302} {\bibfield  {journal}
  {\bibinfo  {journal} {Phys. Rev. B}\ }\textbf {\bibinfo {volume} {100}},\
  \bibinfo {pages} {014302} (\bibinfo {year} {2019})}\BibitemShut {NoStop}%
\bibitem [{\citenamefont {Zangeneh-Nejad}\ and\ \citenamefont
  {Fleury}(2019)}]{Zangeneh2019}%
  \BibitemOpen
  \bibfield  {author} {\bibinfo {author} {\bibfnamefont {F.}~\bibnamefont
  {Zangeneh-Nejad}}\ and\ \bibinfo {author} {\bibfnamefont {R.}~\bibnamefont
  {Fleury}},\ }\bibfield  {title} {\emph {\bibinfo {title} {{Nonlinear
  Second-Order Topological Insulators}},\ }}\href {\doibase
  10.1103/PhysRevLett.123.053902} {\bibfield  {journal} {\bibinfo  {journal}
  {Phys. Rev. Lett.}\ }\textbf {\bibinfo {volume} {123}},\ \bibinfo {pages}
  {053902} (\bibinfo {year} {2019})}\BibitemShut {NoStop}%
\bibitem [{\citenamefont {Chen}\ \emph {et~al.}(2014)\citenamefont {Chen},
  \citenamefont {Upadhyaya},\ and\ \citenamefont {Vitelli}}]{Chen2014}%
  \BibitemOpen
  \bibfield  {author} {\bibinfo {author} {\bibfnamefont {B.~G.-g.}\
  \bibnamefont {Chen}}, \bibinfo {author} {\bibfnamefont {N.}~\bibnamefont
  {Upadhyaya}}, \ and\ \bibinfo {author} {\bibfnamefont {V.}~\bibnamefont
  {Vitelli}},\ }\bibfield  {title} {\emph {\bibinfo {title} {{Nonlinear
  conduction via solitons in a topological mechanical insulator}},\ }}\href
  {\doibase 10.1073/pnas.1405969111} {\bibfield  {journal} {\bibinfo  {journal}
  {Proc. Natl. Acad. Sci. USA}\ }\textbf {\bibinfo {volume} {111}},\ \bibinfo
  {pages} {13004} (\bibinfo {year} {2014})}\BibitemShut {NoStop}%
\bibitem [{\citenamefont {Hadad}\ \emph {et~al.}(2017)\citenamefont {Hadad},
  \citenamefont {Vitelli},\ and\ \citenamefont {Alu}}]{Hadad2017}%
  \BibitemOpen
  \bibfield  {author} {\bibinfo {author} {\bibfnamefont {Y.}~\bibnamefont
  {Hadad}}, \bibinfo {author} {\bibfnamefont {V.}~\bibnamefont {Vitelli}}, \
  and\ \bibinfo {author} {\bibfnamefont {A.}~\bibnamefont {Alu}},\ }\bibfield
  {title} {\emph {\bibinfo {title} {{Solitons and Propagating Domain Walls in
  Topological Resonator Arrays}},\ }}\href {\doibase
  10.1021/acsphotonics.7b00303} {\bibfield  {journal} {\bibinfo  {journal} {ACS
  Photon.}\ }\textbf {\bibinfo {volume} {4}},\ \bibinfo {pages} {1974}
  (\bibinfo {year} {2017})}\BibitemShut {NoStop}%
\bibitem [{\citenamefont {Poddubny}\ and\ \citenamefont
  {Smirnova}(2018)}]{Poddubny2018}%
  \BibitemOpen
  \bibfield  {author} {\bibinfo {author} {\bibfnamefont {A.~N.}\ \bibnamefont
  {Poddubny}}\ and\ \bibinfo {author} {\bibfnamefont {D.~A.}\ \bibnamefont
  {Smirnova}},\ }\bibfield  {title} {\emph {\bibinfo {title} {{Ring Dirac
  solitons in nonlinear topological systems}},\ }}\href {\doibase
  10.1103/PhysRevA.98.013827} {\bibfield  {journal} {\bibinfo  {journal} {Phys.
  Rev. A}\ }\textbf {\bibinfo {volume} {98}},\ \bibinfo {pages} {013827}
  (\bibinfo {year} {2018})}\BibitemShut {NoStop}%
\bibitem [{\citenamefont {Engelhardt}\ \emph {et~al.}(2017)\citenamefont
  {Engelhardt}, \citenamefont {Benito}, \citenamefont {Platero},\ and\
  \citenamefont {Brandes}}]{Engelhardt2017}%
  \BibitemOpen
  \bibfield  {author} {\bibinfo {author} {\bibfnamefont {G.}~\bibnamefont
  {Engelhardt}}, \bibinfo {author} {\bibfnamefont {M.}~\bibnamefont {Benito}},
  \bibinfo {author} {\bibfnamefont {G.}~\bibnamefont {Platero}}, \ and\
  \bibinfo {author} {\bibfnamefont {T.}~\bibnamefont {Brandes}},\ }\bibfield
  {title} {\emph {\bibinfo {title} {Topologically enforced bifurcations in
  superconducting circuits},\ }}\href {\doibase 10.1103/PhysRevLett.118.197702}
  {\bibfield  {journal} {\bibinfo  {journal} {Phys. Rev. Lett.}\ }\textbf
  {\bibinfo {volume} {118}},\ \bibinfo {pages} {197702} (\bibinfo {year}
  {2017})}\BibitemShut {NoStop}%
\bibitem [{\citenamefont {Gerasimenko}\ \emph {et~al.}(2016)\citenamefont
  {Gerasimenko}, \citenamefont {Tarasinski},\ and\ \citenamefont
  {Beenakker}}]{Gerasimenko2016}%
  \BibitemOpen
  \bibfield  {author} {\bibinfo {author} {\bibfnamefont {Y.}~\bibnamefont
  {Gerasimenko}}, \bibinfo {author} {\bibfnamefont {B.}~\bibnamefont
  {Tarasinski}}, \ and\ \bibinfo {author} {\bibfnamefont {C.~W.}\ \bibnamefont
  {Beenakker}},\ }\bibfield  {title} {\emph {\bibinfo {title}
  {{Attractor-repeller pair of topological zero modes in a nonlinear quantum
  walk}},\ }}\href {\doibase 10.1103/PhysRevA.93.022329} {\bibfield  {journal}
  {\bibinfo  {journal} {Phys. Rev. A}\ }\textbf {\bibinfo {volume} {93}},\
  \bibinfo {pages} {022329} (\bibinfo {year} {2016})}\BibitemShut {NoStop}%
\bibitem [{\citenamefont {Bisianov}\ \emph {et~al.}(2019)\citenamefont
  {Bisianov}, \citenamefont {Wimmer}, \citenamefont {Peschel},\ and\
  \citenamefont {Egorov}}]{Bisianov2019}%
  \BibitemOpen
  \bibfield  {author} {\bibinfo {author} {\bibfnamefont {A.}~\bibnamefont
  {Bisianov}}, \bibinfo {author} {\bibfnamefont {M.}~\bibnamefont {Wimmer}},
  \bibinfo {author} {\bibfnamefont {U.}~\bibnamefont {Peschel}}, \ and\
  \bibinfo {author} {\bibfnamefont {O.~A.}\ \bibnamefont {Egorov}},\ }\bibfield
   {title} {\emph {\bibinfo {title} {{Stability of topologically protected edge
  states in nonlinear fiber loops}},\ }}\href {\doibase
  10.1103/PhysRevA.100.063830} {\bibfield  {journal} {\bibinfo  {journal}
  {Phys. Rev. A}\ }\textbf {\bibinfo {volume} {100}},\ \bibinfo {pages}
  {063830} (\bibinfo {year} {2019})}\BibitemShut {NoStop}%
\bibitem [{\citenamefont {Mochizuki}\ \emph {et~al.}(2020)\citenamefont
  {Mochizuki}, \citenamefont {Kawakami},\ and\ \citenamefont
  {Obuse}}]{Mochizuki2020}%
  \BibitemOpen
  \bibfield  {author} {\bibinfo {author} {\bibfnamefont {K.}~\bibnamefont
  {Mochizuki}}, \bibinfo {author} {\bibfnamefont {N.}~\bibnamefont {Kawakami}},
  \ and\ \bibinfo {author} {\bibfnamefont {H.}~\bibnamefont {Obuse}},\
  }\bibfield  {title} {\emph {\bibinfo {title} {{Stability of topologically
  protected edge states in nonlinear quantum walks: Additional bifurcations
  unique to Floquet systems}},\ }}\href {\doibase 10.1088/1751-8121/ab6514}
  {\bibfield  {journal} {\bibinfo  {journal} {J. Phys. A: Math. Theor.}\
  }\textbf {\bibinfo {volume} {53}},\ \bibinfo {pages} {085702} (\bibinfo
  {year} {2020})}\BibitemShut {NoStop}%
\bibitem [{\citenamefont {Lumer}\ \emph {et~al.}(2016)\citenamefont {Lumer},
  \citenamefont {Rechtsman}, \citenamefont {Plotnik},\ and\ \citenamefont
  {Segev}}]{Lumer2016}%
  \BibitemOpen
  \bibfield  {author} {\bibinfo {author} {\bibfnamefont {Y.}~\bibnamefont
  {Lumer}}, \bibinfo {author} {\bibfnamefont {M.~C.}\ \bibnamefont
  {Rechtsman}}, \bibinfo {author} {\bibfnamefont {Y.}~\bibnamefont {Plotnik}},
  \ and\ \bibinfo {author} {\bibfnamefont {M.}~\bibnamefont {Segev}},\
  }\bibfield  {title} {\emph {\bibinfo {title} {{Instability of bosonic
  topological edge states in the presence of interactions}},\ }}\href {\doibase
  10.1103/PhysRevA.94.021801} {\bibfield  {journal} {\bibinfo  {journal} {Phys.
  Rev. A}\ }\textbf {\bibinfo {volume} {94}},\ \bibinfo {pages} {021801(R)}
  (\bibinfo {year} {2016})}\BibitemShut {NoStop}%
\bibitem [{\citenamefont {Shi}\ \emph {et~al.}(2017)\citenamefont {Shi},
  \citenamefont {Kimble},\ and\ \citenamefont {Cirac}}]{Shi2017}%
  \BibitemOpen
  \bibfield  {author} {\bibinfo {author} {\bibfnamefont {T.}~\bibnamefont
  {Shi}}, \bibinfo {author} {\bibfnamefont {H.~J.}\ \bibnamefont {Kimble}}, \
  and\ \bibinfo {author} {\bibfnamefont {J.~I.}\ \bibnamefont {Cirac}},\
  }\bibfield  {title} {\emph {\bibinfo {title} {Topological phenomena in
  classical optical networks},\ }}\href {\doibase 10.1073/pnas.1708944114}
  {\bibfield  {journal} {\bibinfo  {journal} {Proc. Natl. Acad. Sci. USA}\
  }\textbf {\bibinfo {volume} {114}},\ \bibinfo {pages} {E8967} (\bibinfo
  {year} {2017})}\BibitemShut {NoStop}%
\bibitem [{\citenamefont {Palmero}\ \emph {et~al.}(2020)\citenamefont
  {Palmero}, \citenamefont {English}, \citenamefont {Cuevas-Maraver},\ and\
  \citenamefont {Kevrekidis}}]{Palmero2020}%
  \BibitemOpen
  \bibfield  {author} {\bibinfo {author} {\bibfnamefont {F.}~\bibnamefont
  {Palmero}}, \bibinfo {author} {\bibfnamefont {L.}~\bibnamefont {English}},
  \bibinfo {author} {\bibfnamefont {J.}~\bibnamefont {Cuevas-Maraver}}, \ and\
  \bibinfo {author} {\bibfnamefont {P.}~\bibnamefont {Kevrekidis}},\ }\bibfield
   {title} {\emph {\bibinfo {title} {Nonlinear edge modes in a honeycomb
  electrical lattice near the dirac points},\ }}\href {\doibase
  https://doi.org/10.1016/j.physleta.2020.126664} {\bibfield  {journal}
  {\bibinfo  {journal} {Phys. Lett. A}\ }\textbf {\bibinfo {volume} {384}},\
  \bibinfo {pages} {126664} (\bibinfo {year} {2020})}\BibitemShut {NoStop}%
\bibitem [{\citenamefont {Theocharis}\ \emph {et~al.}(2013)\citenamefont
  {Theocharis}, \citenamefont {Boechler},\ and\ \citenamefont
  {Daraio}}]{Theocharis2013}%
  \BibitemOpen
  \bibfield  {author} {\bibinfo {author} {\bibfnamefont {G.}~\bibnamefont
  {Theocharis}}, \bibinfo {author} {\bibfnamefont {N.}~\bibnamefont
  {Boechler}}, \ and\ \bibinfo {author} {\bibfnamefont {C.}~\bibnamefont
  {Daraio}},\ }\bibinfo {title} {Nonlinear periodic phononic structures and
  granular crystals},\ in\ \href {\doibase 10.1007/978-3-642-31232-8_7} {\emph
  {\bibinfo {booktitle} {Acoustic Metamaterials and Phononic Crystals}}},\
  \bibinfo {editor} {edited by\ \bibinfo {editor} {\bibfnamefont {P.~A.}\
  \bibnamefont {Deymier}}}\ (\bibinfo  {publisher} {Springer Berlin
  Heidelberg},\ \bibinfo {address} {Berlin, Heidelberg},\ \bibinfo {year}
  {2013})\ pp.\ \bibinfo {pages} {217--251}\BibitemShut {NoStop}%
\bibitem [{\citenamefont {Fraternali}\ \emph {et~al.}(2015)\citenamefont
  {Fraternali}, \citenamefont {Carpentieri},\ and\ \citenamefont
  {Amendola}}]{Fraternali2015}%
  \BibitemOpen
  \bibfield  {author} {\bibinfo {author} {\bibfnamefont {F.}~\bibnamefont
  {Fraternali}}, \bibinfo {author} {\bibfnamefont {G.}~\bibnamefont
  {Carpentieri}}, \ and\ \bibinfo {author} {\bibfnamefont {A.}~\bibnamefont
  {Amendola}},\ }\bibfield  {title} {\emph {\bibinfo {title} {{On the
  mechanical modeling of the extreme softening/stiffening response of axially
  loaded tensegrity prisms}},\ }}\href {\doibase 10.1016/j.jmps.2014.10.010}
  {\bibfield  {journal} {\bibinfo  {journal} {J. Mech. Phys. Solids}\ }\textbf
  {\bibinfo {volume} {74}},\ \bibinfo {pages} {136} (\bibinfo {year}
  {2015})}\BibitemShut {NoStop}%
\bibitem [{\citenamefont {Yasuda}\ \emph {et~al.}(2019)\citenamefont {Yasuda},
  \citenamefont {Miyazawa}, \citenamefont {Charalampidis}, \citenamefont
  {Chong}, \citenamefont {Kevrekidis},\ and\ \citenamefont
  {Yang}}]{Yasuda2019}%
  \BibitemOpen
  \bibfield  {author} {\bibinfo {author} {\bibfnamefont {H.}~\bibnamefont
  {Yasuda}}, \bibinfo {author} {\bibfnamefont {Y.}~\bibnamefont {Miyazawa}},
  \bibinfo {author} {\bibfnamefont {E.~G.}\ \bibnamefont {Charalampidis}},
  \bibinfo {author} {\bibfnamefont {C.}~\bibnamefont {Chong}}, \bibinfo
  {author} {\bibfnamefont {P.~G.}\ \bibnamefont {Kevrekidis}}, \ and\ \bibinfo
  {author} {\bibfnamefont {J.}~\bibnamefont {Yang}},\ }\bibfield  {title}
  {\emph {\bibinfo {title} {Origami-based impact mitigation via rarefaction
  solitary wave creation},\ }}\href
  {https://advances.sciencemag.org/content/5/5/eaau2835} {\bibfield  {journal}
  {\bibinfo  {journal} {Sci. Adv.}\ }\textbf {\bibinfo {volume} {5}},\ \bibinfo
  {pages} {eaau2835} (\bibinfo {year} {2019})}\BibitemShut {NoStop}%
\bibitem [{\citenamefont {Deng}\ \emph {et~al.}(2020)\citenamefont {Deng},
  \citenamefont {Chen}, \citenamefont {Wei}, \citenamefont {Tournat},\ and\
  \citenamefont {Bertoldi}}]{Deng2020}%
  \BibitemOpen
  \bibfield  {author} {\bibinfo {author} {\bibfnamefont {B.}~\bibnamefont
  {Deng}}, \bibinfo {author} {\bibfnamefont {L.}~\bibnamefont {Chen}}, \bibinfo
  {author} {\bibfnamefont {D.}~\bibnamefont {Wei}}, \bibinfo {author}
  {\bibfnamefont {V.}~\bibnamefont {Tournat}}, \ and\ \bibinfo {author}
  {\bibfnamefont {K.}~\bibnamefont {Bertoldi}},\ }\bibfield  {title} {\emph
  {\bibinfo {title} {Pulse-driven robot: Motion via solitary waves},\ }}\href
  {\doibase 10.1126/sciadv.aaz1166} {\bibfield  {journal} {\bibinfo  {journal}
  {Sci. Adv.}\ }\textbf {\bibinfo {volume} {6}},\ \bibinfo {pages} {eaaz1166}
  (\bibinfo {year} {2020})}\BibitemShut {NoStop}%
\bibitem [{\citenamefont {Jutte}\ and\ \citenamefont {Kota}(2008)}]{Jutte2008}%
  \BibitemOpen
  \bibfield  {author} {\bibinfo {author} {\bibfnamefont {C.~V.}\ \bibnamefont
  {Jutte}}\ and\ \bibinfo {author} {\bibfnamefont {S.}~\bibnamefont {Kota}},\
  }\bibfield  {title} {\emph {\bibinfo {title} {{Design of Nonlinear Springs
  for Prescribed Load-Displacement Functions}},\ }}\href {\doibase
  10.1115/1.2936928} {\bibfield  {journal} {\bibinfo  {journal} {J. Mech.
  Des.}\ }\textbf {\bibinfo {volume} {130}},\ \bibinfo {pages} {081403}
  (\bibinfo {year} {2008})}\BibitemShut {NoStop}%
\bibitem [{\citenamefont {Wang}\ \emph {et~al.}(2014)\citenamefont {Wang},
  \citenamefont {Sigmund},\ and\ \citenamefont {Jensen}}]{Wang2014}%
  \BibitemOpen
  \bibfield  {author} {\bibinfo {author} {\bibfnamefont {F.}~\bibnamefont
  {Wang}}, \bibinfo {author} {\bibfnamefont {O.}~\bibnamefont {Sigmund}}, \
  and\ \bibinfo {author} {\bibfnamefont {J.}~\bibnamefont {Jensen}},\
  }\bibfield  {title} {\emph {\bibinfo {title} {Design of materials with
  prescribed nonlinear properties},\ }}\href {\doibase
  https://doi.org/10.1016/j.jmps.2014.05.003} {\bibfield  {journal} {\bibinfo
  {journal} {J. Mech. Phys. Solids}\ }\textbf {\bibinfo {volume} {69}},\
  \bibinfo {pages} {156 } (\bibinfo {year} {2014})}\BibitemShut {NoStop}%
\bibitem [{\citenamefont {Clausen}\ \emph {et~al.}(2015)\citenamefont
  {Clausen}, \citenamefont {Wang}, \citenamefont {Jensen}, \citenamefont
  {Sigmund},\ and\ \citenamefont {Lewis}}]{Clausen2015}%
  \BibitemOpen
  \bibfield  {author} {\bibinfo {author} {\bibfnamefont {A.}~\bibnamefont
  {Clausen}}, \bibinfo {author} {\bibfnamefont {F.}~\bibnamefont {Wang}},
  \bibinfo {author} {\bibfnamefont {J.~S.}\ \bibnamefont {Jensen}}, \bibinfo
  {author} {\bibfnamefont {O.}~\bibnamefont {Sigmund}}, \ and\ \bibinfo
  {author} {\bibfnamefont {J.~A.}\ \bibnamefont {Lewis}},\ }\bibfield  {title}
  {\emph {\bibinfo {title} {Topology optimized architectures with programmable
  poisson's ratio over large deformations},\ }}\href {\doibase
  10.1002/adma.201502485} {\bibfield  {journal} {\bibinfo  {journal} {Adv.
  Mater.}\ }\textbf {\bibinfo {volume} {27}},\ \bibinfo {pages} {5523}
  (\bibinfo {year} {2015})}\BibitemShut {NoStop}%
\bibitem [{\citenamefont {Hussein}\ \emph {et~al.}(2014)\citenamefont
  {Hussein}, \citenamefont {Leamy},\ and\ \citenamefont
  {Ruzzene}}]{Hussein2014}%
  \BibitemOpen
  \bibfield  {author} {\bibinfo {author} {\bibfnamefont {M.~I.}\ \bibnamefont
  {Hussein}}, \bibinfo {author} {\bibfnamefont {M.~J.}\ \bibnamefont {Leamy}},
  \ and\ \bibinfo {author} {\bibfnamefont {M.}~\bibnamefont {Ruzzene}},\
  }\bibfield  {title} {\emph {\bibinfo {title} {{Dynamics of Phononic Materials
  and Structures: Historical Origins, Recent Progress, and Future Outlook}},\
  }}\href {\doibase 10.1115/1.4026911} {\bibfield  {journal} {\bibinfo
  {journal} {Appl. Mech. Rev.}\ }\textbf {\bibinfo {volume} {66}},\ \bibinfo
  {pages} {040802} (\bibinfo {year} {2014})}\BibitemShut {NoStop}%
\bibitem [{\citenamefont {Cummer}\ \emph {et~al.}(2016)\citenamefont {Cummer},
  \citenamefont {Christensen},\ and\ \citenamefont {Al{\`{u}}}}]{Cummer2016}%
  \BibitemOpen
  \bibfield  {author} {\bibinfo {author} {\bibfnamefont {S.~A.}\ \bibnamefont
  {Cummer}}, \bibinfo {author} {\bibfnamefont {J.}~\bibnamefont {Christensen}},
  \ and\ \bibinfo {author} {\bibfnamefont {A.}~\bibnamefont {Al{\`{u}}}},\
  }\bibfield  {title} {\emph {\bibinfo {title} {{Controlling sound with
  acoustic metamaterials}},\ }}\href {\doibase 10.1038/natrevmats.2016.1}
  {\bibfield  {journal} {\bibinfo  {journal} {Nat. Rev. Mater.}\ }\textbf
  {\bibinfo {volume} {1}},\ \bibinfo {pages} {16001} (\bibinfo {year}
  {2016})}\BibitemShut {NoStop}%
\bibitem [{\citenamefont {Bertoldi}\ \emph {et~al.}(2017)\citenamefont
  {Bertoldi}, \citenamefont {Vitelli}, \citenamefont {Christensen},\ and\
  \citenamefont {van Hecke}}]{Bertoldi2017}%
  \BibitemOpen
  \bibfield  {author} {\bibinfo {author} {\bibfnamefont {K.}~\bibnamefont
  {Bertoldi}}, \bibinfo {author} {\bibfnamefont {V.}~\bibnamefont {Vitelli}},
  \bibinfo {author} {\bibfnamefont {J.}~\bibnamefont {Christensen}}, \ and\
  \bibinfo {author} {\bibfnamefont {M.}~\bibnamefont {van Hecke}},\ }\bibfield
  {title} {\emph {\bibinfo {title} {{Flexible mechanical metamaterials}},\
  }}\href {https://www.nature.com/articles/natrevmats201766} {\bibfield
  {journal} {\bibinfo  {journal} {Nat. Rev. Mater.}\ }\textbf {\bibinfo
  {volume} {2}},\ \bibinfo {pages} {17066} (\bibinfo {year}
  {2017})}\BibitemShut {NoStop}%
\bibitem [{\citenamefont {Su}\ \emph {et~al.}(1979)\citenamefont {Su},
  \citenamefont {Schrieffer},\ and\ \citenamefont {Heeger}}]{Su1979}%
  \BibitemOpen
  \bibfield  {author} {\bibinfo {author} {\bibfnamefont {W.~P.}\ \bibnamefont
  {Su}}, \bibinfo {author} {\bibfnamefont {J.~R.}\ \bibnamefont {Schrieffer}},
  \ and\ \bibinfo {author} {\bibfnamefont {A.~J.}\ \bibnamefont {Heeger}},\
  }\bibfield  {title} {\emph {\bibinfo {title} {{Solitons in Polyacetylene}},\
  }}\href {\doibase 10.1103/PhysRevLett.42.1698} {\bibfield  {journal}
  {\bibinfo  {journal} {Phys. Rev. Lett.}\ }\textbf {\bibinfo {volume} {42}},\
  \bibinfo {pages} {1698} (\bibinfo {year} {1979})}\BibitemShut {NoStop}%
\bibitem [{\citenamefont {Prodan}\ and\ \citenamefont
  {Prodan}(2009)}]{Prodan2009}%
  \BibitemOpen
  \bibfield  {author} {\bibinfo {author} {\bibfnamefont {E.}~\bibnamefont
  {Prodan}}\ and\ \bibinfo {author} {\bibfnamefont {C.}~\bibnamefont
  {Prodan}},\ }\bibfield  {title} {\emph {\bibinfo {title} {Topological phonon
  modes and their role in dynamic instability of microtubules},\ }}\href
  {\doibase 10.1103/PhysRevLett.103.248101} {\bibfield  {journal} {\bibinfo
  {journal} {Phys. Rev. Lett.}\ }\textbf {\bibinfo {volume} {103}},\ \bibinfo
  {pages} {248101} (\bibinfo {year} {2009})}\BibitemShut {NoStop}%
\bibitem [{\citenamefont {Chaunsali}\ \emph {et~al.}(2017)\citenamefont
  {Chaunsali}, \citenamefont {Kim}, \citenamefont {Thakkar}, \citenamefont
  {Kevrekidis},\ and\ \citenamefont {Yang}}]{Chaunsali2017}%
  \BibitemOpen
  \bibfield  {author} {\bibinfo {author} {\bibfnamefont {R.}~\bibnamefont
  {Chaunsali}}, \bibinfo {author} {\bibfnamefont {E.}~\bibnamefont {Kim}},
  \bibinfo {author} {\bibfnamefont {A.}~\bibnamefont {Thakkar}}, \bibinfo
  {author} {\bibfnamefont {P.~G.}\ \bibnamefont {Kevrekidis}}, \ and\ \bibinfo
  {author} {\bibfnamefont {J.}~\bibnamefont {Yang}},\ }\bibfield  {title}
  {\emph {\bibinfo {title} {{Demonstrating an In Situ Topological Band
  Transition in Cylindrical Granular Chains}},\ }}\href {\doibase
  10.1103/PhysRevLett.119.024301} {\bibfield  {journal} {\bibinfo  {journal}
  {Phys. Rev. Lett.}\ }\textbf {\bibinfo {volume} {119}},\ \bibinfo {pages}
  {024301} (\bibinfo {year} {2017})}\BibitemShut {NoStop}%
\bibitem [{\citenamefont {Marín}\ and\ \citenamefont
  {Aubry}(1998)}]{Marin1998}%
  \BibitemOpen
  \bibfield  {author} {\bibinfo {author} {\bibfnamefont {J.}~\bibnamefont
  {Marín}}\ and\ \bibinfo {author} {\bibfnamefont {S.}~\bibnamefont {Aubry}},\
  }\bibfield  {title} {\emph {\bibinfo {title} {Finite size effects on
  instabilities of discrete breathers},\ }}\href {\doibase
  https://doi.org/10.1016/S0167-2789(98)00077-3} {\bibfield  {journal}
  {\bibinfo  {journal} {Physica D}\ }\textbf {\bibinfo {volume} {119}},\
  \bibinfo {pages} {163 } (\bibinfo {year} {1998})}\BibitemShut {NoStop}%
\bibitem [{\citenamefont {Aubry}(2006)}]{Aubry2006}%
  \BibitemOpen
  \bibfield  {author} {\bibinfo {author} {\bibfnamefont {S.}~\bibnamefont
  {Aubry}},\ }\bibfield  {title} {\emph {\bibinfo {title} {{Discrete Breathers:
  Localization and transfer of energy in discrete Hamiltonian nonlinear
  systems}},\ }}\href {\doibase 10.1016/j.physd.2005.12.020} {\bibfield
  {journal} {\bibinfo  {journal} {Physica D}\ }\textbf {\bibinfo {volume}
  {216}},\ \bibinfo {pages} {1} (\bibinfo {year} {2006})}\BibitemShut {NoStop}%
\bibitem [{\citenamefont {Flach}\ and\ \citenamefont
  {Gorbach}(2008)}]{Flach2008}%
  \BibitemOpen
  \bibfield  {author} {\bibinfo {author} {\bibfnamefont {S.}~\bibnamefont
  {Flach}}\ and\ \bibinfo {author} {\bibfnamefont {A.~V.}\ \bibnamefont
  {Gorbach}},\ }\bibfield  {title} {\emph {\bibinfo {title} {{Discrete
  breathers — Advances in theory and applications}},\ }}\href {\doibase
  10.1016/j.physrep.2008.05.002} {\bibfield  {journal} {\bibinfo  {journal}
  {Physics Reports}\ }\textbf {\bibinfo {volume} {467}},\ \bibinfo {pages} {1}
  (\bibinfo {year} {2008})}\BibitemShut {NoStop}%
\bibitem [{\citenamefont {Gershgorin}\ \emph {et~al.}(2005)\citenamefont
  {Gershgorin}, \citenamefont {Lvov},\ and\ \citenamefont
  {Cai}}]{Gershgorin2005}%
  \BibitemOpen
  \bibfield  {author} {\bibinfo {author} {\bibfnamefont {B.}~\bibnamefont
  {Gershgorin}}, \bibinfo {author} {\bibfnamefont {Y.~V.}\ \bibnamefont
  {Lvov}}, \ and\ \bibinfo {author} {\bibfnamefont {D.}~\bibnamefont {Cai}},\
  }\bibfield  {title} {\emph {\bibinfo {title} {Renormalized waves and discrete
  breathers in $\ensuremath{\beta}$-fermi-pasta-ulam chains},\ }}\href
  {\doibase 10.1103/PhysRevLett.95.264302} {\bibfield  {journal} {\bibinfo
  {journal} {Phys. Rev. Lett.}\ }\textbf {\bibinfo {volume} {95}},\ \bibinfo
  {pages} {264302} (\bibinfo {year} {2005})}\BibitemShut {NoStop}%
\bibitem [{\citenamefont {Jiang}\ \emph {et~al.}(2014)\citenamefont {Jiang},
  \citenamefont {Lu}, \citenamefont {Zhou},\ and\ \citenamefont
  {Cai}}]{Jiang2014}%
  \BibitemOpen
  \bibfield  {author} {\bibinfo {author} {\bibfnamefont {S.-x.~W.}\
  \bibnamefont {Jiang}}, \bibinfo {author} {\bibfnamefont {H.-h.}\ \bibnamefont
  {Lu}}, \bibinfo {author} {\bibfnamefont {D.}~\bibnamefont {Zhou}}, \ and\
  \bibinfo {author} {\bibfnamefont {D.}~\bibnamefont {Cai}},\ }\bibfield
  {title} {\emph {\bibinfo {title} {Renormalized dispersion relations of
  $\ensuremath{\beta}$-fermi-pasta-ulam chains in equilibrium and
  nonequilibrium states},\ }}\href {\doibase 10.1103/PhysRevE.90.032925}
  {\bibfield  {journal} {\bibinfo  {journal} {Phys. Rev. E}\ }\textbf {\bibinfo
  {volume} {90}},\ \bibinfo {pages} {032925} (\bibinfo {year}
  {2014})}\BibitemShut {NoStop}%
\bibitem [{\citenamefont {Chung}\ \emph {et~al.}(2020)\citenamefont {Chung},
  \citenamefont {Cheon},\ and\ \citenamefont {Qin}}]{Chung2020}%
  \BibitemOpen
  \bibfield  {author} {\bibinfo {author} {\bibfnamefont {M.}~\bibnamefont
  {Chung}}, \bibinfo {author} {\bibfnamefont {Y.-L.}\ \bibnamefont {Cheon}}, \
  and\ \bibinfo {author} {\bibfnamefont {H.}~\bibnamefont {Qin}},\ }\bibfield
  {title} {\emph {\bibinfo {title} {Linear beam stability in periodic focusing
  systems: Krein signature and band structure},\ }}\href {\doibase
  https://doi.org/10.1016/j.nima.2020.163708} {\bibfield  {journal} {\bibinfo
  {journal} {Nuclear Inst. and Methods in Physics Research, A}\ }\textbf
  {\bibinfo {volume} {962}},\ \bibinfo {pages} {163708} (\bibinfo {year}
  {2020})}\BibitemShut {NoStop}%
\bibitem [{\citenamefont {Marín}\ \emph {et~al.}(1998)\citenamefont {Marín},
  \citenamefont {Eilbeck},\ and\ \citenamefont {Russell}}]{Marin1998b}%
  \BibitemOpen
  \bibfield  {author} {\bibinfo {author} {\bibfnamefont {J.}~\bibnamefont
  {Marín}}, \bibinfo {author} {\bibfnamefont {J.}~\bibnamefont {Eilbeck}}, \
  and\ \bibinfo {author} {\bibfnamefont {F.}~\bibnamefont {Russell}},\
  }\bibfield  {title} {\emph {\bibinfo {title} {Localized moving breathers in a
  2d hexagonal lattice},\ }}\href {\doibase
  https://doi.org/10.1016/S0375-9601(98)00577-5} {\bibfield  {journal}
  {\bibinfo  {journal} {Phys. Lett. A}\ }\textbf {\bibinfo {volume} {248}},\
  \bibinfo {pages} {225 } (\bibinfo {year} {1998})}\BibitemShut {NoStop}%
\end{thebibliography}%

\end{document}